\documentclass[zpreprint,zbstnp]{zeus_paper}

\usepackage[english]{babel}

\newcommand{\ZcoosysA}{%
The ZEUS coordinate system is a right-handed Cartesian system, with the $Z$
axis pointing in the proton beam direction, referred to as the ``forward
direction'', and the $X$ axis pointing left towards the center of HERA.
The coordinate origin is at the nominal interaction point.\xspace}

\newcommand{\ZcoosysfnA}{\footnote{\ZcoosysA}}

\newcommand{\Zdetdesc}{%
A detailed description of the ZEUS detector can be found 
elsewhere~\cite{zeus:1993:bluebook}. A brief outline of the 
components that are most relevant for this analysis is given
below.\xspace}
\newcommand{\Zctddesc}[1]{%
Charged particles are tracked in the central tracking detector (CTD)~\citeCTD,
which operates in a magnetic field of $1.43\Tesla$ provided by a thin 
superconducting coil. The CTD consists of 72~cylindrical drift chamber 
layers, organised in 9~superlayers covering the polar-angle#1 region 
\mbox{$15^\circ<\theta<164^\circ$}. The transverse-momentum resolution for
full-length tracks is $\sigma(p_T)/p_T=0.0058p_T\oplus0.0065\oplus0.0014/p_T$,
with $p_T$ in $\Gev$.}
\newcommand{\Zcaldesc}{%
The high-resolution uranium--scintillator calorimeter (CAL)~\citeCAL consists 
of three parts: the forward (FCAL), the barrel (BCAL) and the rear (RCAL)
calorimeters. Each part is subdivided transversely into towers and
longitudinally into one electromagnetic section (EMC) and either one (in RCAL)
or two (in BCAL and FCAL) hadronic sections (HAC). The smallest subdivision of
the calorimeter is called a cell.  The CAL energy resolutions, as measured under
test-beam conditions, are $\sigma(E)/E=0.18/\sqrt{E}$ for electrons and
$\sigma(E)/E=0.35/\sqrt{E}$ for hadrons ($E$ in $\Gev$).}

\chardef\usc=95
\chardef\til=126
\catcode`\@=11 
\DeclareRobustCommand\xdotspace{\futurelet\@let@token\@xdotspace}
\def\@xdotspace{%
  \ifx\@let@token.\else
  \ifx\@let@token\bgroup.\else
  \ifx\@let@token\egroup.\else
  \ifx\@let@token\/.\else
  \ifx\@let@token\ .\else
  \ifx\@let@token~.\else
  \ifx\@let@token!.\else
  \ifx\@let@token,.\else
  \ifx\@let@token:.\else
  \ifx\@let@token;.\else
  \ifx\@let@token?.\else
  \ifx\@let@token/.\else
  \ifx\@let@token'.\else
  \ifx\@let@token).\else
  \ifx\@let@token-.\else
  \ifx\@let@token\@xobeysp.\else
  \ifx\@let@token\space.\else
  \ifx\@let@token\@sptoken.\else
   .\space
   \fi\fi\fi\fi\fi\fi\fi\fi\fi\fi\fi\fi\fi\fi\fi\fi\fi\fi}
\catcode`\@=12 

\newcommand{\stru}[2]{%
   \relax\ifmmode\hbox{\vrule height#1 depth#2 width0pt}%
   \else\vrule height#1 depth#2 width0pt\fi}

\newcommand{\Ronum}[1]{\uppercase\expandafter{\romannumeral#1}}
\newcommand{\ronum}[1]{\expandafter{\romannumeral#1}}
\DeclareRobustCommand{\LaTeXZ}{%
  \LaTeX\kern-.05em4\kern-.1em
  {\raisebox{-0.2ex}{$\scriptstyle\text{ZEUS}$}}\xspace}

\DeclareMathAlphabet{\mathbf}{OT1}{cmr}{bx}{sl}
\newcommand{\eVdist}{\kern-0.06667em}

\newcommand{\Gev}{{\text{Ge}\eVdist\text{V\/}}}

\newcommand{\gev}{{\,\text{Ge}\eVdist\text{V\/}}}

\newcommand{\cm}{\,\text{cm}}

\newcommand{\Tesla}{\,\text{T}}

\newcommand{\slashfrac}[2]{%
  \raisebox{0.5ex}{\ensuremath #1}\kern-0.12em/\kern-0.08em
  \raisebox{-.8ex}{\ensuremath #2}}

\newcommand{\sqr}[3]{%
    {\vcenter{\hrule height.#3ex\hbox{\vrule width.#2ex height#1ex
     \kern#1ex\vrule width.#3ex}\hrule height.#2ex}}}

\catcode`\@=11 
\newcommand{\parenbar}{\mathpalette\p@renb@r}
\def\p@renb@r#1#2{\vbox{%
  \ifx#1\scriptscriptstyle \dimen@.7em\dimen@ii.2em\else
  \ifx#1\scriptstyle \dimen@.8em\dimen@ii.25em\else
  \dimen@1em\dimen@ii.4em\fi\fi \offinterlineskip
  \ialign{\hfill##\hfill\cr
    \vbox{\hrule width\dimen@ii}\cr
    \noalign{\vskip-.3ex}%
    \hbox to\dimen@{$\mathchar300\hfil\mathchar301$}\cr
    \noalign{\vskip-.3ex}%
    $#1#2$\cr}}}
\catcode`\@=12 

\newcommand{\IP}{{\rm I$\kern-0.01667em$P}\xspace}

\mathchardef\qsm=63
\mathchardef\pls=43
\mathchardef\mns=512
\mathchardef\plm=518
\mathchardef\eql=61
\mathchardef\smallleft=300
\mathchardef\smallright=301
\mathchardef\les=316
\mathchardef\gre=318
\mathchardef\leq=532
\mathchardef\grq=533

\catcode`\@=11 
\newcounter{pict@width}
\newcounter{pict@height}
\newlength{\pict@scale}
\newcommand{\psfigadd}[4]{%
\setcounter{pict@width}{1*\ratio{#2+\pict@scale/2}{\pict@scale}}
\setcounter{pict@height}{1*\ratio{#3+\pict@scale/2}{\pict@scale}}
\setlength{\unitlength}{\pict@scale}
\hbox to #2{\hspace{-\fill}\begin{picture}(\thepict@width,\thepict@height)
\put(0,0){\psfig{figure=#1,width=#2,height=#3,clip=}}
\SetScale{0.283466457}
\SetWidth{1.763889}
{#4}
\end{picture}}
}
\newcounter{pict@widthfst}
\newcounter{pict@widthscd}
\newcounter{pict@widthtot}
\newcommand{\psfigaddtwo}[7]{%
\setcounter{pict@widthfst}{1*\ratio{#2+\pict@scale/2}{\pict@scale}}
\setcounter{pict@widthscd}{1*\ratio{#2+#4+\pict@scale/2}{\pict@scale}}
\setcounter{pict@widthtot}{1*\ratio{#2+#4+#6+\pict@scale/2}{\pict@scale}}
\setcounter{pict@height}{1*\ratio{#3+\pict@scale/2}{\pict@scale}}
\setlength{\unitlength}{\pict@scale}
\hbox{\hspace{-\fill}\begin{picture}(\thepict@widthtot,\thepict@height)
\put(0,0){\psfig{figure=#1,width=#2,height=#3,clip=}}
\put(\thepict@widthscd,0){\psfig{figure=#5,width=#6,height=#3,clip=}}
\SetScale{0.283466457}
\SetWidth{1.763889}
{#7}
\end{picture}}
}
\newcommand{\psfigror}[4]{%
\setcounter{pict@width}{1*\ratio{#2+\pict@scale/2}{\pict@scale}}
\setcounter{pict@height}{1*\ratio{#3+\pict@scale/2}{\pict@scale}}
\setlength{\unitlength}{\pict@scale}
\hbox{\begin{picture}(\thepict@width,\thepict@height)
\put(0,\thepict@height){\psfig{figure=#1,width=#3,height=#2,clip=,angle=270}}
\SetScale{0.283466457}
\SetWidth{1.763889}
{#4}
\end{picture}}
}
\newcommand{\psfigrol}[4]{%
\setcounter{pict@width}{1*\ratio{#2+\pict@scale/2}{\pict@scale}}
\setcounter{pict@height}{1*\ratio{#3+\pict@scale/2}{\pict@scale}}
\setlength{\unitlength}{\pict@scale}
\hbox{\begin{picture}(\thepict@width,\thepict@height)
\put(0,0){\psfig{figure=#1,width=#3,height=#2,clip=,angle=90}}
\SetScale{0.283466457}
\SetWidth{1.763889}
{#4}
\end{picture}}
}
\catcode`\@=12 
\newlength\listtextwidth

\catcode`\@=11 
\newlength{\@tabfninsert}
\newlength{\@tabfnwidth}
\newcommand{\tabfootnote}[2]{%
  \setlength{\@tabfninsert}{0.8em}
  \setlength{\@tabfnwidth}{\textwidth}
  \addtolength{\@tabfnwidth}{-\@tabfninsert}
  \addtolength{\@tabfnwidth}{-0.4em}
  \noindent\makebox[\@tabfninsert][r]{\footnotesize$^{#1}$\hfil}\hfill%
  \parbox[t]{\@tabfnwidth}{\footnotesize #2\hfill}}
\catcode`\@=12 

\def\citeCTD{{\cite{%
nim:a279:290,*npps:b32:181,*nim:a338:254%
}}\xspace}
\def\citeCAL{{\cite{%
nim:a309:77,*nim:a309:101,*nim:a321:356,*nim:a336:23%
}}\xspace}

\begin{document}

\prepnum{{DESY--05--132}}

\title{
Inclusive jet cross sections and \\
dijet correlations in $D^{*\pm}$ photoproduction \\ at HERA
}

\author{ZEUS Collaboration}

\date{July 2005}

\abstract{
Inclusive jet cross sections in photoproduction for events containing a $D^*$ meson 
have been measured with the ZEUS detector at HERA using an integrated luminosity 
of $78.6\,{\rm pb}^{-1}$. The events were required to have a virtuality of the 
incoming photon, $Q^2$, of less than  1\,GeV$^2$, and a photon-proton centre-of-mass 
energy in the range $130<W_{\gamma p}<280\,{\rm GeV}$. The measurements are compared 
with next-to-leading-order (NLO) QCD calculations. Good agreement is found with the NLO
calculations over most of the measured kinematic region. Requiring a second jet in the 
event allowed a more detailed comparison with QCD calculations. The measured dijet cross 
sections are also compared to Monte Carlo (MC) models which incorporate 
leading-order matrix elements followed by parton showers and hadronisation. The NLO QCD 
predictions are in general agreement with the data although differences have been 
isolated to regions where contributions from higher orders are expected to be significant. 
The MC models give a better description than the NLO predictions of the shape of the 
measured cross sections.
}

\makezeustitle

\def\3{\ss}

\pagenumbering{Roman}

\begin{center}
{                      \Large  The ZEUS Collaboration              }
\end{center}

{
  S.~Chekanov,                                                                                     
  M.~Derrick,                                                                                      
  S.~Magill,                                                                                       
  S.~Miglioranzi$^{   1}$,                                                                         
  B.~Musgrave,                                                                                     
  \mbox{J.~Repond},                                                                                
  R.~Yoshida\\                                                                                     
 {\it Argonne National Laboratory, Argonne, Illinois 60439-4815}, USA~$^{n}$                       
\par \filbreak                                                                                     
  M.C.K.~Mattingly \\                                                                              
 {\it Andrews University, Berrien Springs, Michigan 49104-0380}, USA                               
\par \filbreak                                                                                     
  N.~Pavel, A.G.~Yag\"ues Molina \\                                                                
  {\it Institut f\"ur Physik der Humboldt-Universit\"at zu Berlin,                                 
           Berlin, Germany}
% ...
\newpage                                                                                           
$^{\    1}$ also affiliated with University College London, UK \\                                  
$^{\    2}$ now at Siemens VDO/Sensorik, Weissensberg \\                                           
% ...
                                                           %
                                                           %
\newpage   
                                                           %
                                                           %
\begin{tabular}[h]{rp{14cm}}                                                                       
$^{a}$ &  supported by the Natural Sciences and Engineering Research Council of Canada (NSERC) \\  
$^{b}$ &  supported by the German Federal Ministry for Education and Research (BMBF), under        
          contract numbers HZ1GUA 2, HZ1GUB 0, HZ1PDA 5, HZ1VFA 5\\                                
$^{c}$ &  supported in part by the MINERVA Gesellschaft f\"ur Forschung GmbH, the Israel Science   
          Foundation (grant no. 293/02-11.2), the U.S.-Israel Binational Science Foundation and    
          the Benozyio Center for High Energy Physics\\                                            
$^{d}$ &  supported by the German-Israeli Foundation and the Israel Science Foundation\\           
$^{e}$ &  supported by the Italian National Institute for Nuclear Physics (INFN) \\                
$^{f}$ &  supported by the Japanese Ministry of Education, Culture, Sports, Science and Technology 
          (MEXT) and its grants for Scientific Research\\                                          
$^{g}$ &  supported by the Korean Ministry of Education and Korea Science and Engineering          
          Foundation\\                                                                             
$^{h}$ &  supported by the Netherlands Foundation for Research on Matter (FOM)\\                   
$^{i}$ &  supported by the Polish State Committee for Scientific Research, grant no.               
          620/E-77/SPB/DESY/P-03/DZ 117/2003-2005 and grant no. 1P03B07427/2004-2006\\             
$^{j}$ &  partially supported by the German Federal Ministry for Education and Research (BMBF)\\   
$^{k}$ &  supported by RF Presidential grant N 1685.2003.2 for the leading scientific schools and  
          by the Russian Ministry of Education and Science through its grant for Scientific        
          Research on High Energy Physics\\                                                        
$^{l}$ &  supported by the Spanish Ministry of Education and Science through funds provided by     
          CICYT\\                                                                                  
$^{m}$ &  supported by the Particle Physics and Astronomy Research Council, UK\\                   
$^{n}$ &  supported by the US Department of Energy\\                                               
$^{o}$ &  supported by the US National Science Foundation\\                                        
$^{p}$ &  supported by the Polish Ministry of Scientific Research and Information Technology,      
          grant no. 112/E-356/SPUB/DESY/P-03/DZ 116/2003-2005 and 1 P03B 065 27\\                  
$^{q}$ &  supported by FNRS and its associated funds (IISN and FRIA) and by an Inter-University    
          Attraction Poles Programme subsidised by the Belgian Federal Science Policy Office\\     
\end{tabular}

:}

\newpage

\newcommand{\DELPHIJJ}{$\Delta \phi^{\rm jj}$}
\newcommand{\PTJJ}{$(p_{T}^{\rm jj})^2$}
\newcommand{\MJJ}{$m^{\rm jj}$}
\newcommand{\XGOBS}{$x_{\gamma}^{\rm obs}$}

{
\section{Introduction}
\label{sec-int}

Charm and/or jet production in $ep$ collisions should be accurately calculable 
in perturbative Quantum Chromodynamics (pQCD) since the mass of the heavy quark, 
$m_{Q}$, and the transverse energy of the jet, $E_{T}^{\rm jet}$, provide hard 
scales. In photoproduction, where a quasi-real photon, emitted from the incoming 
lepton, collides with a parton from the incoming proton, such events can be 
classified into two types of process in leading-order (LO) QCD. In direct 
processes, the photon couples as a point-like object in the hard scatter. In 
resolved processes, the photon acts as a source of incoming partons with only a 
fraction of its momentum participating in the hard scatter.

Measurements of the $D^*$ photoproduction cross section~\cite{epj:c6:67} as 
functions of the transverse momentum, $p_{T}^{D^{*}}$, and the pseudorapidity, 
$\eta^{D^{*}}$, show that the predictions from next-to-leading-order (NLO) QCD 
are too low for $p_{T}^{D^{*}} >$~3\,GeV and $\eta^{D^*}> 0$. Part of this 
deficit may be due to hadronisation effects. The predictions for jet production 
accompanied by a $D^*$ meson should have smaller uncertainties from these 
hadronisation effects. Furthermore jets can be measured in a wider pseudorapidity 
range than $D^*$ mesons due to the larger acceptance of the calorimeter compared 
to the central tracker.

A dijet sample of $D^{*}$ photoproduction can also be used to study higher-order 
QCD topologies~\cite{epj:c6:67,pl:b565:87}. 
In the present paper, previously unmeasured correlations between the two jets of 
highest transverse energy, namely the difference in azimuthal angle,~\DELPHIJJ, 
and the squared transverse momentum of the dijet system,~\PTJJ, which are 
particularly sensitive to higher-order topologies, are presented. For the LO 
$2\rightarrow2$ process, the two jets are produced back-to-back with 
$\Delta\phi^{\rm jj}=\pi$ and $(p_{T}^{\rm jj})^2=0$. Large deviations from these 
values may come from higher-order QCD effects. The accuracy of the theoretical 
description of these effects is tested.

Calculations performed to NLO in QCD are available with two different treatments 
for charm. In the fixed-order, or ``massive'', scheme~\cite{pl:b348:633,*np:b454:3}, 
$u$, $d$ and $s$ are the only active flavours in the structure functions of the 
proton and photon; charm and beauty are produced only in the hard scatter. This 
scheme is expected to work well in regions where the transverse momentum of the 
outgoing $c$ quark is of the order of the quark mass. At higher transverse momenta, 
the resummed or ``massless'' scheme~\cite{zfp:c76:677,*zfp:c76:689,pr:d58:014014} 
should be applicable. In this scheme, charm and beauty are regarded as active flavours 
(massless partons) in the structure functions of the proton and photon and are 
fragmented from massless partons into massive hadrons after the hard process. 

In this paper, photoproduction of charm is studied by tagging a $D^{*\pm}$ meson 
and reconstructing at least one jet in the final state. The measurement is performed 
in the following kinematic region: photon virtuality, $Q^{2}<1\gev^{2}$; photon-proton 
centre-of-mass energy, $130<W_{\gamma p}<280~{\rm GeV}$; $p_{T}^{D^{*}}>3\gev$; 
$|\eta^{D^{*}}|<1.5$; jet transverse energy, $E_{T}^{\rm jet}>6\gev$; and jet 
pseudorapidity, $-1.5<\eta^{\rm jet}<2.4$. Differential cross sections as a function 
of $E_{T}^{\rm jet}$ and $\eta^{\rm jet}$ have been measured. Jets are divided 
into two categories: jets of the first category are associated with the 
$D^{*}$ meson ($D^{*}$-tagged jet), while jets of the second category are not 
matched to a $D^*$ meson (untagged jet). The inclusive, $D^*$-tagged and untagged 
jet cross sections are compared to the massive NLO QCD predictions. A comparison 
to the massless calculation is only available for the untagged jet cross 
sections~\cite{pr:d70:094035}.

A sub-sample having at least two jets with $E_{T}^{\rm jet1}>7\gev$ and 
$E_{T}^{\rm jet2}>6\gev$ is used to measure the correlations between the two 
highest $E_{T}^{\rm jet}$ jets: the fraction of the photon momentum participating in 
dijet production, $x_{\gamma}^{\rm obs}$~\cite{pl:b348:665}; $\Delta\phi^{\rm jj}$; 
$(p_{T}^{\rm jj})^2$; and the dijet invariant mass, $M^{\rm jj}$.
Differential dijet cross sections as a function of these variables have been measured 
for the direct-enriched ($x_{\gamma}^{\rm obs} > 0.75$) and resolved-enriched 
($x_{\gamma}^{\rm obs} < 0.75$) kinematic regions and compared to massive NLO QCD predictions 
and Monte Carlo (MC) models.


\section{Experimental set-up}
\label{sec-exp}

The analysis was performed with data taken from 1998 to 2000, when HERA collided 
electrons or positrons\footnote{Hereafter, both electrons and positrons are referred 
to as electrons, unless explicitly stated otherwise.} with energy $E_e =$ 27.5\,GeV 
with protons of energy $E_p =$ 920\,GeV resulting in a centre-of-mass energy of 
318\,GeV. The results are based on an integrated luminosity of $78.6~{\rm pb}^{-1}$  
of $ep$ collision data taken by the ZEUS detector.
\Zdetdesc

\Zctddesc\ZcoosysfnA

\Zcaldesc

The luminosity was measured from the rate of the bremsstrahlung process 
$ep~\rightarrow~e\gamma p$, where the photon was measured in a lead--scintillator
calorimeter~\cite{desy-92-066,*zfp:c63:391,*acpp:b32:2025} placed in the HERA 
tunnel at $Z=-107~{\rm m}$.


\section{Event reconstruction and selection}

A three-level trigger system was used to select events 
online~\cite{zeus:1993:bluebook,proc:chep:1992:222}. At the first- and second-level 
triggers, general characteristics of photoproduction events were required and 
background due to beam-gas interactions rejected. At the third level, a $D^*$ 
candidate was reconstructed. 

In the offline analysis, the hadronic final state and jets were reconstructed using 
a combination of track and calorimeter information that optimises the resolution of 
reconstructed kinematic variables~\cite{briskin:phd:1998}. The selected tracks and 
calorimeter clusters are referred to as Energy Flow Objects (EFOs). To select 
photoproduction events, the following criteria were used: 

\begin{itemize}

\item the event vertex was required to be within $50\cm$ of the nominal vertex 
      position in the longitudinal direction;

\item deep inelastic scattering (DIS) events with a scattered electron candidate in 
      the CAL were removed~\cite{pl:b322:287}. To keep the events where a pion was 
      misidentified as a scattered electron, events where 
      \mbox{$y_e=1-\frac{E_{e^\prime}}{2E_e}(1-\cos\theta_{e^\prime})>$  0.7} were 
      retained; $E_{e^\prime}$ and $\theta_{e^\prime}$ are the energy and polar angle, 
      respectively, of the scattered electron candidate;

\item the requirement $130 <W_{\rm JB}<280$\,GeV was imposed, where 
      $W_{\rm JB}=\sqrt{4E_{p}E_{e}y_{\rm JB}}$ and $y_{\rm JB}$ is the estimator of 
      the inelasticity, $y$, measured from the EFOs according to the Jacquet-Blondel 
      method~\cite{proc:epfacility:1979:391}. Here, $W_{\rm JB}$ was corrected, using MC 
      simulation, for the energy losses of EFOs in inactive material in front of the 
      CAL. The upper cut removed DIS events where the scattered electron was not identified 
      and which, therefore, have a value of $y_{\rm JB}$ close to 1. The lower cut removed 
      proton beam-gas events which have a low value of $y_{\rm JB}$. 

\end{itemize}

The cuts on $y_e$ and $W_{\rm JB}$ restricted the range of the virtuality of the 
exchanged photon to $Q^2$ less than about 1~GeV$^2$, with a median value of about 
\mbox{3 $\times$ 10$^{-4}$~${\rm GeV^2}$}. 

\subsection{$D^{*}$ reconstruction}
\label{sec:dstar-rec}

The $D^*$ mesons were identified using the decay channel 
$D^{*+}\to D^0\pi^{+}_s$ with the subsequent decay $D^0\to K^-\pi^+$ and the 
corresponding antiparticle decay, where $\pi^+_s$ refers to a low-momentum 
(``slow'') pion accompanying the $D^0$.

Charged tracks measured by the CTD and assigned to the primary event vertex were 
selected. The transverse momentum was required to be greater than 0.12~GeV. Each 
track was required to reach at least the third superlayer of the CTD. These 
restrictions ensured that the track acceptance and momentum resolution were 
high. Tracks in the CTD with opposite charges and transverse momenta 
$p_T > 0.4\gev$ were combined in pairs to form $D^0$ candidates. The tracks were 
alternately assigned the masses of a kaon and a pion and the invariant mass of 
the pair, $M_{K\pi}$, was evaluated. Each additional track, with charge opposite to 
that of the kaon track, was assigned the pion mass and combined with the 
$D^0$-meson candidate to form a $D^*$ candidate.  
 
The signal regions for the reconstructed masses, $M(D^0)$ and 
$\Delta M=(M_{K\pi\pi_s} - M_{K\pi})$, were \mbox{$1.80 < M(D^0)<1.92$ GeV} and 
\mbox{$0.143 < \Delta M < 0.148$ GeV}, respectively. For background determination, 
$D^0$ candidates with wrong-sign combinations, in which both 
tracks forming the $D^0$ candidates have the same charge and the third track has 
the opposite charge, were also retained. The same kinematic restrictions were 
applied as for those $D^0$ candidates with correct-charge combinations. The 
normalisation factor of the wrong-charge sample was determined as the ratio of 
events with correct-charge combinations to wrong-charge combinations in the 
region \mbox{$0.15<\Delta M<0.17$ GeV}.  

The kinematic region for $D^*$ candidates was $p_T^{D^*} > 3$ GeV and 
\mbox{$|\eta^{D^*}|<1.5$}. Figure~\ref{fig-deltam} shows $\Delta M$ for 
the selection of a $D^*$ meson with a jet (see Section~\ref{sec:jet}). 
The fit to the distribution has the form

$$
F =p_1 \cdot 
\exp \left( -0.5\cdot x^{1+\frac{1}{1+0.5x}} \right) + 
p_4\cdot (\Delta M - m_\pi)^{p_5} ,
$$

where $x=|(\Delta M-p_2)/p_3|$, $p_1-p_5$ are free parameters and $m_\pi$ is the 
pion mass. The ``modified'' Gaussian described both data and MC distributions well. 
The fit gives a peak at \mbox{$145.467 \pm 0.015$(stat.) MeV} 
to be compared to the PDG value of \mbox{$145.421\pm 0.010$ MeV~\cite{pl:b592:1}}. 
The difference is due to systematic effects which are too small to be relevant for 
this analysis. The fitted width of \mbox{$0.61 \pm 0.02$ MeV} is consistent with the 
experimental resolution. 

The number of $D^*$ mesons was determined from candidates reconstructed in both signal 
regions and after the subtraction of the background estimated from the wrong-charge 
sample; this gave $4891\pm 113$ $D^*$ mesons. This procedure was used throughout the 
paper with the number of $D^*$ mesons obtained from the fit used as a systematic check. 

\subsection{Jet reconstruction}
\label{sec:jet}

Jets were reconstructed using EFOs as input to the the $k_T$ cluster 
algorithm~\cite{np:b406:187} in its longitudinally invariant inclusive 
mode~\cite{pr:d48:3160}. The transverse energy of the jet was corrected for energy 
losses in inactive material in front of the CAL, where the correction factors were 
determined in bins of $E_{T}^{\rm jet}$ and $\eta^{\rm jet}$ from MC simulation. 
These corrections were between 5\% and 12\%. 

For the inclusive jet cross sections, jets with $E_{T}^{\rm jet}>6~\gev$ and 
$-1.5<\eta^{\rm jet}<2.4$ were selected, and events containing at least one such jet 
were used for further analysis. For the dijet analysis, events were required to have 
at least two jets with $-1.5<\eta^{\rm jet}<2.4$ and $E_{T}^{\rm jet}>6\gev$, while 
the highest $E_{T}^{\rm jet}$ jet was required in addition to satisfy 
$E_{T}^{\rm jet}>7\gev$. The asymmetric jet transverse-energy cut assures that the 
NLO calculation is not infrared sensitive~\cite{np:b507:315}. After the dijet 
selection, $1692 \pm 70$ $D^*$ mesons remained.

Cross sections are given separately for $D^*$-tagged and untagged jets. $D^*$-tagged 
jets are defined as jets in which a $D^*$ (in the kinematic region defined in 
Section~\ref{sec:dstar-rec}) was clustered into the jet at the hadron level by the $k_T$ 
algorithm. All the other jets are called untagged jets. These two classes of jets can be 
distinguished experimentally by cutting on the distance $\Delta R(D^{*},{\rm jet})$ 
$=\sqrt{ (\eta^{D^{*}}-\eta^{\rm jet})^2-(\phi^{D^{*}}-\phi^{\rm jet})^2}$ between the 
$D^*$ and the jet, where $\phi^{D^*}$ and $\phi^{\rm jet}$ are the azimuthal angles of 
the $D^*$ meson and jet, respectively. Figure~\ref{fig-deltar} shows the distance of the 
$D^*$ to all jets in the event for data and MC. The peak at $\Delta R(D^{*},{\rm jet})=0$ 
is due to $D^*$-tagged jets as indicated by the $\Delta R(D^{*},{\rm jet})$ distribution 
for $D^*$-tagged jets for the MC hadron-level predictions also shown in 
Fig.~\ref{fig-deltar}. The broad peak at $\Delta R(D^{*},{\rm jet}) \sim 3$ is due to the 
untagged jets. A cut at $\Delta R(D^{*},{\rm jet}) < 0.6$ was used to distinguish 
experimentally tagged and untagged jets. For the inclusive jet sample, 83\% of $D^*$ mesons 
were matched to a jet, and for the dijet sub-sample 94\% of $D^*$ mesons were matched to a 
jet.


\section{Monte Carlo models}
\label{sec:mc}

The MC programmes {\sc Herwig 6.301}~\cite{hep-ph-9912396,hep-ph-0107071} and 
{\sc Pythia 6.156}~\cite{cpc:135:238}, which implement LO matrix 
elements followed by parton showers and hadronisation, were used to model the 
final state. The {\sc Herwig} and {\sc Pythia} generators differ in the details of 
the implementation of the leading-logarithmic parton-shower models. They also use 
different hadronisation models: {\sc Herwig} uses the cluster~\cite{np:b238:492} 
model and {\sc Pythia} uses the Lund string~\cite{prep:97:31} model. Direct and 
resolved events were generated separately and in proportion to the cross sections 
predicted by the MC programme. The relative fraction of charm and beauty events was 
also generated in proportion to the cross sections predicted by the MC programme. 
Events were generated using CTEQ5L~\cite{epj:c12:375} and \mbox{GRV-G LO} 
parton density functions (PDF) for the proton and the photon, respectively. 
The $c$-quark and $b$-quark masses were set to $m_c=1.5~{\rm GeV}$ and 
$m_b=4.75~{\rm GeV}$, respectively.

The MC programmes were used both to compare with the dijet cross sections, which 
are particularly sensitive to the parton-shower models and for calculation of 
the acceptance and effects of detector response (see Section~\ref{sec:corr_syst}). 
For all generated events, the ZEUS detector response was simulated in detail using 
a programme based on {\sc Geant}~3.13~\cite{tech:cern-dd-ee-84-1}.


\section{NLO QCD calculations}

There are two NLO QCD calculations available to calculate jet cross sections
in charm photoproduction: the massive calculation by Frixione et al. 
(FMNR)~\cite{pl:b348:633,*np:b454:3} and the massless calculation by Heinrich 
and Kniehl~\cite{pr:d70:094035}. 

\subsection{Massive calculation}

In the massive calculation, the PDF sets used were CTEQ5M1~\cite{epj:c12:375} for 
the proton and AFG-HO~\cite{zfp:c64:621} for the photon. The renormalisation scale, $\mu_R$, 
and factorisation scale, $\mu_F$, were set to 
$\mu = \mu_{R} = \mu_F = m_{T} = \sqrt{\langle (p_T^c)^2 \rangle + m_{c}^2}$, where $\langle (p_T^c)^2 \rangle$ is the 
average squared transverse momentum of the two charm quarks and $m_{c}=1.5\gev$.  The fragmentation 
of the charm quark into a $D^{*}$ 
meson was described by rescaling the $c$-quark momentum using the Peterson fragmentation 
function~\cite{pr:d27:105} with $\epsilon=0.035 \pm 0.002$ which is taken from an NLO 
fit to ARGUS data~\cite{np:b565:245}. The fraction of charm quarks hadronising into a 
$D^*$ meson was set to 0.235~\cite{hep-ex-9912064}. The $k_{T}$ algorithm was applied 
to the outgoing partons in the final state of the NLO programme.

The dependence of the NLO prediction on different photon PDFs, $\mu_R$, $\mu_F$ and 
$m_c$ was evaluated by repeating the calculation using different sets of parameters. 
The upper (lower) bound of the NLO QCD prediction was estimated by setting 
$\mu_R = m_T/2$ and $m_{c}=1.3\gev$ ($\mu_R = 2 m_T$ and $m_{c}=1.7\gev$).

An NLO prediction of $D^*$ production for beauty is not available so this contribution 
was estimated using a combination of the $B$ hadron cross section at NLO and $B$ decays 
in {\sc Pythia}. The $p_T$ 
distributions of the two stable $B$ hadrons produced in the 
{\sc Pythia} MC programme were reweighted to the distribution in the NLO calculation. In the 
NLO calculation, the $b$-quark mass, $m_b$, was set to 4.75\,GeV, 
$\mu = m_{T} = \sqrt{\langle (p_T^b)^2 \rangle + m_b^2}$ and $\epsilon = 0.0035$~\cite{np:b623:247}. 
The branching $b \to D^*$ was set to the value measured by the OPAL 
Collaboration~\cite{epj:c1:439}. The 
upper (lower) bound of the NLO QCD prediction was estimated by setting $\mu_R = m_T/2$ and 
$m_b=4.5\gev$ ($\mu_R = 2 m_T$ and $m_{c}=5.0\gev$). The contribution from beauty 
production for the inclusive jet distribution, as predicted by NLO+{\sc Pythia}, is about 
2\% at low $E_{T}^{\rm jet}$ and increases to 8\% at high $E_{T}^{\rm jet}$. For each 
cross section, this beauty contribution and its uncertainty were added linearly to the 
corresponding massive $D^*$ prediction from charm quarks. 

\subsection{Massless calculation}

In the massless calculation, AFG04~\cite{hep-ph-0503259} for the photon PDF and 
MRST03~\cite{npps:b79:105,*epj:c23:73} for the proton PDF were used.
The number of flavours was set to five. The $D^{*}$ fragmentation function and fraction of 
beauty and charm hadronising into a $D^*$ meson are derived~\cite{pr:d58:014014,pr:d70:094035} 
from a 
fit to data from LEP; the function is assumed to be applicable to HERA as it is 
derived using the factorisation theorem in QCD. The central prediction uses 
$\mu_{R} = m_{T}^\prime = \sqrt{(p_T^{D^*})^2 + m_{c}^2}$, where $m_c=1.5\gev$. 
The fragmentation factorisation~\cite{pr:d70:094035} scale, $M_{F}$, and $\mu_F$ 
were set to $M_F = \mu_{F}= 2m_{T}^\prime$. 

The uncertainty was estimated by changing the scales to $\mu_{R} = m_{T}^\prime/2$ 
and $\mu_{F} = M_F = 4m_{T}^\prime$ for the upper bound, and $\mu_{R} = 2m_{T}^\prime$ 
and $\mu_{F} = M_F = m_{T}^\prime$ for the lower bound. The photon PDF, GRV-HO, was 
used, and the uncertainty was found to be less than half that from the variation in 
scale~\cite{pr:d70:094035}. The difference between MRST01 and MRST03 proton PDFs was found 
to be negligible for the distributions considered. The size of the beauty contribution was 
estimated by suppressing the final-state fragmentation of a $b$ quark to a $D^*$ meson. 
This reduces the cross-section prediction by about 3\% at low $E_{T}^{\rm jet}$ and 15\% at 
high $E_{T}^{\rm jet}$. Due to theoretical limitations, the predictions for the massless 
scheme can only be calculated for the untagged-jet distributions.

\subsection{Hadronisation correction}

As the NLO calculations produce final-state partons, the effects of hadronisation 
were considered when comparing the predictions with the data. The NLO QCD jet 
predictions were corrected using a bin-by-bin procedure according to 
$d\sigma = d\sigma^{\rm NLO} \cdot C_{\rm had}$, where $d\sigma^{\rm NLO}$ is the 
cross section for parton jets in the final state of the NLO calculation. The 
hadronisation correction factor was defined as the ratio of the jet cross sections 
after and before the hadronisation process,
$C_{\rm had}=d\sigma_{\rm MC}^{\rm hadrons}/d\sigma_{\rm MC}^{\rm partons}$. Here, 
parton-level cross sections were obtained using partons after the initial- and final-state 
showering of the MC simulations described in Section~\ref{sec:mc}. Distributions 
at the parton level in the MC programmes were checked to be similar to those calculated 
using the NLO programme, assuring the validity of using a bin-by-bin correction. The 
value of $C_{\rm had}$ was taken as the mean of the ratios obtained using the 
{\sc Herwig} and {\sc Pythia} predictions. The uncertainty on this value was estimated 
as half the difference between the values obtained using the two models. These 
uncertainties were added in quadrature to the other uncertainties of the NLO 
calculations. The values of $C_{\rm had}$ applied to the NLO predictions are given in 
Tables~\ref{tbl:inclusive_et}--\ref{tbl:inclusive_mjj}.


\section{Data correction and systematic uncertainties}
\label{sec:corr_syst}

The data were corrected, using the MC models described in Section~\ref{sec:mc}, for the 
detector acceptance and the selection efficiencies to obtain differential cross 
sections for the process $ep \rightarrow e^\prime + D^{*} + {\rm jet(s)} + X$. The definition 
of the cross section includes events with a $D^{*}$ containing primary $c$ quarks or 
those from  
$b$-quark decays. The data were initially compared to the MC simulation in shape and 
found generally to agree well for all the kinematic quantities. Since {\sc Herwig} 
gives a better overall description of the data than {\sc Pythia}, it was chosen as 
the primary MC model to correct the data. The cross section for a given 
observable $Y$ was determined using 

\begin{equation}
\frac {d\sigma}{dY} = 
\frac {N } {A \cdot \mathcal {L} \cdot B \cdot \Delta Y},  
\nonumber 
\end{equation}

where $N$ is either the number of jets for the inclusive jet cross sections or the number of 
$D^*$ mesons for the dijet cross sections in a bin of size $\Delta Y$. The acceptance, $A$, 
takes into account migrations and efficiencies for that bin, and $\mathcal {L}$ is the 
integrated luminosity. The product, $B$, of the appropriate branching ratios for the $D^*$ 
and $D^0$ was set to $(2.57\pm 0.06)\%$~\cite{pl:b592:1}. 

The systematic uncertainties of the measured cross sections were determined by changing 
the selection cuts or the analysis procedure in turn and repeating the extraction of the 
cross sections. The uncertainties are described in more detail 
elsewhere~\cite{thesis:kohno:2004}. The following systematic studies have been carried out 
(the resulting uncertainty on the total cross section is given in parentheses):

\begin{itemize}

\item varying the values of the selection cuts by the experimental resolutions in 
      the corresponding quantity ($^{+2.7}_{-2.3}\%$);

\item varying the efficiencies of the CAL first-level trigger ($+4.1\%$);

\item the acceptance was recalculated by re-weighting the prediction from the 
      {\sc Herwig} MC simulation in $p_{T}^{D^{*}}/E_{T}^{\rm jet}$ to reproduce 
      the distribution of this variable in the data ($+5.0$\%);

\item an additional contribution to the  uncertainty from the modelling of the 
      hadronisation process was 
      estimated by using {\sc Pythia} instead of {\sc Herwig} ($+0.4$\%);

\item the effect of the uncertainty of the beauty cross section on the acceptance 
      correction was taken into account by increasing the beauty contribution by
      a factor two ($<\pm 1$\%);

\item varying the procedure to extract the $D^*$ signal ($^{+1.3}_{-2.9}$\%);
 
\item varying by $\pm3\%$ the jet energy scale in the CAL ($^{+2.7}_{-2.5}$\%).

\end{itemize}

All systematic uncertainties were added in quadrature to obtain the total systematic 
uncertainty, except for the jet energy-scale uncertainty, as this has a large 
correlation between bins, and is shown separately in all figures. In most bins of the 
differential cross sections, the total systematic uncertainty is comparable to the 
statistical errors. In addition, an overall normalisation uncertainty of $2\%$ from 
the luminosity determination is included in neither the figures nor the tables.


\section{Inclusive jet cross sections}
\label{sec:incl}

Inclusive jet cross sections with a $D^{*}$ in the final state have been measured 
as a function of $E_{T}^{\rm jet}$ and $\eta^{\rm jet}$ in the following kinematic 
region: 

\begin{itemize}

  \item {$Q^2 < 1~\gev^2$ and $130<W_{\gamma p}<280\gev$;} 

  \item {$p_{T}^{D^{*}}>3\gev$ and $|\eta^{D^{*}}|<1.5$;}

  \item {$E_{T}^{\rm jet}>6\gev$ and $-1.5<\eta^{\rm jet}<2.4$;}

  \item {a $D^{*}$-jet match is required only where explicitly stated.}

\end{itemize}

At the hadron level, the $D^*$ meson is used as input into the jet finder therefore 
allowing a $D^*$-tagged jet to be identified unambiguously.

The cross-sections $d\sigma/dE_{T}^{\rm jet}$ in bins of $\eta^{\rm jet}$ are shown in 
Fig.~\ref{fig-et} and given in Table~\ref{tbl:inclusive_et} for all jets and in 
Fig.~\ref{fig-et_dstarother} and Table~\ref{tbl:inclusive_et_dstarother} for the whole 
$\eta^{\rm jet}$ range for $D^*$-tagged and untagged jets. The distributions tend to fall less 
steeply with increasing $\eta^{\rm jet}$. The cross sections $d\sigma/d\eta^{\rm jet}$ 
in different regions of $E_{T}^{\rm jet}$ are shown in Fig.~\ref{fig-eta} and given in 
Table~\ref{tbl:inclusive_eta} for all jets and in Fig.~\ref{fig-eta_dstarother} and 
Tables~\ref{tbl:inclusive_eta_dstar} and~\ref{tbl:inclusive_eta_other} for $D^*$-tagged and untagged jets. Due to the 
requirement $|\eta^{D^{*}}|<1.5$, the $D^*$-tagged jet is centred around 
$\eta^{\rm jet} = 0$ and 
falls off rapidly at large $\eta^{\rm jet}$. The advantage of reconstructing jets 
is observed in the untagged-jet distribution where a significant cross section 
is measured up to $\eta^{\rm jet} =$~2.4. This is due to the larger acceptance of the 
CAL compared to the CTD.

The massive calculation is compared to all measured cross sections whereas the massless 
calculation is compared only to the untagged-jet distributions. The normalisation 
of the data for all distributions is well described by the upper limit of both NLO 
QCD predictions. The shape of the data is well described by the NLO QCD predictions. 
The inclusion of hadronisation corrections which shift the distributions towards 
forward $\eta^{\rm jet}$ improves the description of the data. Some difference in shape 
is observed for the upper bound of the massless prediction compared to the untagged-jet 
distributions for $E_T^{\rm jet} >$~9\,GeV (see Fig.~\ref{fig-eta_dstarother}). 

Measurements of $d\sigma/dE_{T}^{\rm jet}$ and $d\sigma/d\eta^{\rm jet}$ in bins of 
$p_T^{D^*}$ are given in Tables~\ref{tbl:inclusive_et_ptds} 
and~\ref{tbl:inclusive_eta_ptds} and shown in Fig.~\ref{fig-inptbins} compared to the 
massive NLO prediction. The NLO prediction gives a poor description of the normalisation 
of the data at lowest $p_T^{D^*}$. However, the normalisation of the NLO 
prediction agrees with the data in the two regions of higher $p_T^{D^*}$. In all 
regions, the shape of the data is reasonably well described by the NLO prediction. 
Similar conclusions on the normalisation were seen for inclusive $D^*$ 
measurements~\cite{epj:c6:67}. However, the difference in shape observed as a function 
of $\eta^{D^*}$ in the inclusive measurement is not seen here as a function of 
$\eta^{\rm jet}$. The $b$-quark contribution is largest in the region of low $p_T^{D^*}$.

In order to be sensitive to higher-order effects, and to distinguish between 
direct-enriched and resolved-enriched regions, the variable 
$x_{\gamma}^{\rm obs}(D^{*},{\rm jet})$ was constructed~\cite{pr:d70:094035},
in an analogous way to the `traditional' $x_{\gamma}^{\rm obs}$~\cite{pl:b348:665}. 
Using the $D^*$ meson and the untagged jet of highest 
$E_T^{\rm jet}$, the quantity $x_{\gamma}^{\rm obs}(D^{*},{\rm jet})$ is given by:  

\begin{equation}
x_{\gamma}^{\rm obs}(D^*,{\rm jet}) = \frac{p_T^{D^*}e^{-\eta^{D^*}} 
                                + E_T^{(\rm untagged~jet)}e^{-\eta^{(\rm untagged~jet)}}}
                                       {2yE_{e}}~.
\label{xgammadstarother}
\end{equation}

This variable has the advantage of being calculable in the massless scheme. In 
addition it takes advantage of increased statistics by requiring only one jet of high 
$E_{T}^{\rm jet}$. In Fig.~\ref{fig-xgamma_dsjet} the measured cross-section 
$d\sigma/dx_{\gamma}^{\rm obs}(D^*,{\rm jet})$, given in 
Table~\ref{tbl:inclusive_xgamma_dsjet}, is compared to both the massless and massive 
predictions. The upper bound of the massive prediction gives a good description of the 
data; the description of the massless prediction is somewhat worse. The {\sc Herwig} MC 
model gives a poor description whilst {\sc Pythia} gives a reasonable description of 
the shape of the data distribution. Both MC programmes underestimate the normalisation 
of the data.


\section{Dijet cross sections}

Dijet correlations are particularly sensitive to higher-order effects and therefore 
suitable to test the limitations of fixed-order perturbative QCD calculations. Events 
containing a $D^*$ meson were required to have at 
least two jets with \mbox{$E_{T}^{\rm jet1}>7\gev$}, $E_{T}^{\rm jet2}>6\gev$ and 
$-1.5 < \eta^{\rm jet1,2} < 2.4$. The $Q^2$, $W_{\gamma p}$, $p_T^{D^*}$ and 
$\eta^{D^*}$ requirements were the same as for the inclusive jet cross 
section. 

The dijet variables measured were reconstructed from the two highest 
$E_{T}^{\rm jet}$ jets as:

\begin{equation}
x_{\gamma}^{\rm obs} = \frac{E_T^{\rm jet1}e^{-\eta^{\rm jet1}} 
                           + E_T^{\rm jet2}e^{-\eta^{\rm jet2}}}
			    {2yE_{e}} ~,
\label{xgammaeq}
\end{equation}

\begin{equation}
\Delta\phi^{\rm jj} = |\phi^{\rm jet1}-\phi^{\rm jet2}|~,
\label{eqdphijj}
\end{equation}

\begin{equation}
\left( p_{T}^{\rm jj} \right)^2 = \left(p_x^{\rm jet1}+p_x^{\rm jet2}\right)^2 
                                + \left(p_y^{\rm jet1}+p_y^{\rm jet2}\right)^2~,
\label{eqptjj}
\end{equation}

\begin{equation}
M^{\rm jj} = \sqrt{2 E_T^{\rm jet1} E_{T}^{\rm jet2}
                  \left[ \cosh(\eta^{\rm jet1} - \eta^{\rm jet2}) 
                       - \cos(\phi^{\rm jet1} - \phi^{\rm jet2})\right]}~.
\label{eqmjj}
\end{equation}


Table~\ref{tbl:inclusive_xgamma} gives, and Fig.~\ref{fig-xgamma}a shows, the dijet 
cross section as a function of $x_{\gamma}^{\rm obs}$, which is reasonably well described 
by the massive calculation. In Tables~\ref{tbl:inclusive_dphi}-\ref{tbl:inclusive_mjj} 
and Figs.~\ref{fig-xgamma}b-d, the cross 
sections as a function of $\Delta\phi^{\rm jj}$, $(p_{T}^{\rm jj})^2$ and $M^{\rm jj}$ 
are also shown. For \DELPHIJJ~there is agreement between data and the NLO prediction 
at large angular separation, but at smaller \DELPHIJJ~values the NLO prediction 
underestimates the data. This is correlated with the agreement and disagreement 
at low and high \PTJJ~ values, respectively. The distribution in dijet invariant 
mass is described well by the upper limit of the NLO prediction, as was the case for the 
inclusive jet cross sections in Section~\ref{sec:incl}.

Cross sections as a function of $M^{\rm jj}$, \DELPHIJJ~ and \PTJJ~ are shown in 
Figs.~\ref{fig-mjj}--\ref{fig-ptjj} and given in 
Tables~\ref{tbl:inclusive_dphi}-\ref{tbl:inclusive_mjj} separately for direct-enriched 
($x_\gamma^{\rm obs}>0.75$) and resolved-enriched ($x_\gamma^{\rm obs}<0.75$) 
samples. The data are compared to massive NLO QCD predictions and expectations 
from MC models. 

The cross-section $d\sigma/dM^{\rm jj}$ in Fig.~\ref{fig-mjj} is described well by 
the NLO prediction and both MC models, {\sc Herwig} and {\sc Pythia}, for both regions 
in $x_\gamma^{\rm obs}$. 

The cross-sections $d\sigma/d\Delta\phi^{\rm jj}$ (see Fig.~\ref{fig-dphi}) and 
$d\sigma/d(p_{T}^{\rm jj})^2$ (see Fig.~\ref{fig-ptjj}) are reasonably well reproduced 
by the NLO prediction for $x_\gamma^{\rm obs}>0.75$ although the data exhibit a 
somewhat harder distribution. For $x_\gamma^{\rm obs}<0.75$, the data exhibit a harder 
spectrum than for $x_\gamma^{\rm obs}>0.75$. The NLO prediction of the cross section 
for $x_\gamma^{\rm obs}<0.75$ has a significantly softer distribution compared to the data,  
both as a function of \DELPHIJJ~ and \PTJJ. The low-$x_\gamma^{\rm obs}$ region is more 
sensitive to higher-order topologies not present in the massive NLO calculation. The 
predictions from the {\sc Pythia} MC reproduce neither the shape nor the normalisation 
of the data for low and high $x_\gamma^{\rm obs}$. However, the predictions from the 
{\sc Herwig} MC give an excellent description of the shapes of all distributions, although 
the normalisation is underestimated by a factor of 2.5. The fact that a MC programme 
incorporating parton showers can successfully describe the data whereas the NLO QCD 
prediction cannot indicates that the QCD  
calculation requires higher orders. Matching of parton showers with NLO calculations 
such as in the MC@NLO programme~\cite{jhep:0206:029,*jhep:0308:007}, which is not 
currently available for the processes studied here, should improve the description of the 
data.


\section{Conclusions}

Differential inclusive jet cross sections for events containing a $D^*$ meson have been 
measured with the ZEUS detector in the kinematic region $Q^{2}<1~{\rm GeV}^{2}$, 
$130<W_{\gamma p}<280~{\rm GeV}$, \mbox{$p_{T}^{D^{*}}>3\gev$}, $|\eta^{D^{*}}|<1.5$, 
$E_{T}^{\rm jet} > 6~{\rm GeV}$ and $-1.5<\eta^{\rm jet}<2.4$. The measurements are 
compared to NLO QCD predictions in the massive and massless schemes. With the addition 
of hadronisation corrections, the upper limit of both theoretical predictions show 
similar trends and reasonable agreement with all measured cross sections. Dijet correlation 
cross-sections $d\sigma/dM^{\rm jj}$ and $d\sigma/dx_{\gamma}^{\rm obs}$ are well described 
by the massive NLO QCD prediction, although again the data tends to agree better 
with the upper bound of the NLO calculation. In contrast, the cross-sections 
$d\sigma/d\Delta\phi^{\rm jj}$ and $d\sigma/d(p_{T}^{\rm jj})^2$ show a large deviation from 
the massive NLO QCD prediction at low $\Delta\phi^{\rm jj}$ and high $(p_{T}^{\rm jj})^2$.
This discrepancy is enhanced for the resolved-enriched ($x_{\gamma}^{\rm obs} < 0.75$) 
sample. These regions are expected to be particularly sensitive to higher-order 
effects. The {\sc Herwig} MC model which incorporates leading-order matrix elements  
followed by parton showers and hadronisation describes the shape of the measurements 
well. This indicates that for the precise description of charm dijet photoproduction, 
higher-order calculations or the implementation of additional parton showers in 
current NLO calculations are needed.

\section*{Acknowledgements}

We thank the DESY Directorate for their strong support and encouragement. The 
remarkable achievements of the HERA machine group were essential for the successful 
completion of this work. The design, construction and installation of the ZEUS 
detector have been made possible by the effort of many people who are not listed as 
authors. We also thank G.~Heinrich and B.~Kniehl for providing their calculation.
}

{
\providecommand{\etal}{et al.\xspace}
\providecommand{\coll}{Collab.\xspace}
\catcode`\@=11
\def\@bibitem#1{%
\ifmc@bstsupport
  \mc@iftail{#1}%
    {;\newline\ignorespaces}%
    {\ifmc@first\else.\fi\orig@bibitem{#1}}
  \mc@firstfalse
\else
  \mc@iftail{#1}%
    {\ignorespaces}%
    {\orig@bibitem{#1}}%
\fi}%
\catcode`\@=12

}

{
\begin{table}[htbp]
  \begin{center}
    \begin{tabular}{lr|lr|cccccc}
\hline
\multicolumn{2}{c|}{$\eta^{\rm jet}$ range} & \multicolumn{2}{c|}{$E_{T}^{\rm jet}$ range} & $d\sigma/dE_{T}^{\rm jet}$ & $\pm$ stat. & $\pm$ syst. & $\pm$ E-scale &
(nb/GeV) \\
\hline
-1.5 & 2.4 & 6.0 & 9.0 & 1.583 & $\pm$0.039 & $^{+0.145}_{-0.079}$ & $^{+0.072}_{-0.060}$ & \\
\multicolumn{2}{c|}{} & 9.0 & 13.0 & 0.367 & $\pm$0.018 & $^{+0.038}_{-0.017}$ & $^{+0.025}_{-0.023}$ & \\
\multicolumn{2}{c|}{} & 13.0 & 18.0 & 0.0775 & $\pm$0.0092 & $^{+0.0147}_{-0.0073}$ & $^{+0.0097}_{-0.0074}$ & \\
\multicolumn{2}{c|}{} & 18.0 & 25.0 & 0.0269 & $\pm$0.0046 & $^{+0.0053}_{-0.0055}$ & $^{+0.0032}_{-0.0026}$ & \\
\hline
-1.5 & -0.5 & 6.0 & 9.0 & 0.321 & $\pm$0.017 & $^{+0.032}_{-0.010}$ & $^{+0.025}_{-0.018}$ & \\
\multicolumn{2}{c|}{} & 9.0 & 13.0 & 0.0408 & $\pm$0.0062 & $^{+0.0053}_{-0.0039}$ & $^{+0.0044}_{-0.0044}$ & \\
\multicolumn{2}{c|}{} & 13.0 & 18.0 & 0.0088 & $\pm$0.0033 & $^{+0.0013}_{-0.0055}$ & $^{+0.0020}_{-0.0022}$ & \\
\multicolumn{2}{c|}{} & 18.0 & 25.0 & 0.00 & $\pm$0.00 & $^{+0.00}_{0.00}$ & $^{+0.00}_{0.00}$ & \\
\hline
-0.5 & 0.5 & 6.0 & 9.0 & 0.669 & $\pm$0.023 & $^{+0.064}_{-0.033}$ & $^{+0.027}_{-0.023}$ & \\
\multicolumn{2}{c|}{} & 9.0 & 13.0 & 0.159 & $\pm$0.011 & $^{+0.017}_{-0.010}$ & $^{+0.011}_{-0.010}$ & \\
\multicolumn{2}{c|}{} & 13.0 & 18.0 & 0.0315 & $\pm$0.0050 & $^{+0.0056}_{-0.0030}$ & $^{+0.0042}_{-0.0037}$ & \\
\multicolumn{2}{c|}{} & 18.0 & 25.0 & 0.0089 & $\pm$0.0025 & $^{+0.0032}_{-0.0012}$ & $^{+0.0012}_{-0.0010}$ & \\
\hline
0.5 & 1.5 & 6.0 & 9.0 & 0.446 & $\pm$0.022 & $^{+0.048}_{-0.024}$ & $^{+0.016}_{-0.013}$ & \\
\multicolumn{2}{c|}{} & 9.0 & 13.0 & 0.122 & $\pm$0.011 & $^{+0.015}_{-0.010}$ & $^{+0.007}_{-0.006}$ & \\
\multicolumn{2}{c|}{} & 13.0 & 18.0 & 0.0306 & $\pm$0.0063 & $^{+0.0110}_{-0.0032}$ & $^{+0.0034}_{-0.0020}$ & \\
\multicolumn{2}{c|}{} & 18.0 & 25.0 & 0.0126 & $\pm$0.0033 & $^{+0.0032}_{-0.0047}$ & $^{+0.0013}_{-0.0014}$ & \\
\hline
1.5 & 2.4 & 6.0 & 9.0 & 0.151 & $\pm$0.017 & $^{+0.015}_{-0.026}$ & $^{+0.004}_{-0.006}$ & \\
\multicolumn{2}{c|}{} & 9.0 & 13.0 & 0.0462 & $\pm$0.0072 & $^{+0.0075}_{-0.0058}$ & $^{+0.0024}_{-0.0024}$ & \\
\multicolumn{2}{c|}{} & 13.0 & 18.0 & 0.0095 & $\pm$0.0041 & $^{+0.0055}_{-0.0047}$ & $^{+0.0013}_{-0.0008}$ & \\
\multicolumn{2}{c|}{} & 18.0 & 25.0 & 0.0053 & $\pm$0.0019 & $^{+0.0013}_{-0.0015}$ & $^{+0.0007}_{0.0002}$ & \\
\hline
    \end{tabular}
    \caption{The cross section $d\sigma / d E_{T}^{\rm jet}$ for events containing 
             at least one $D^*$ meson in different regions of $\eta^{\rm jet}$. 
             The statistical (stat.), systematic (syst.) and energy scale (E-scale) 
             uncertainties are shown separately.}
    \label{tbl:inclusive_et}
  \end{center}
\end{table}
\begin{table}[htbp]
  \begin{center}
    \begin{tabular}{c|lr|ccccc}
\hline
$E_{T}^{\rm jet}$ range & \multicolumn{2}{c|}{$\eta^{\rm jet}$ range} & $d\sigma/d\eta^{\rm jet}$ & $\pm$ stat. & $\pm$ syst. & $\pm$ E-scale & (nb) \\
\hline
$E_{T}^{\rm jet}>6$ & -1.5 & -1.0 & 0.548 & $\pm$0.069 & $^{+0.086}_{-0.156}$ & $^{+0.065}_{-0.053}$ & \\
 & -1.0 & -0.5 & 1.745 & $\pm$0.094 & $^{+0.210}_{-0.059}$ & $^{+0.135}_{-0.100}$ & \\
 & -0.5 & 0.0 & 2.79 & $\pm$0.12 & $^{+0.26}_{-0.18}$ & $^{+0.16}_{-0.14}$ & \\
 & 0.0 & 0.5 & 2.90 & $\pm$0.12 & $^{+0.31}_{-0.11}$ & $^{+0.13}_{-0.12}$ & \\
 & 0.5 & 1.0 & 2.30 & $\pm$0.12 & $^{+0.28}_{-0.10}$ & $^{+0.11}_{-0.09}$ & \\
 & 1.0 & 1.5 & 1.85 & $\pm$0.13 & $^{+0.15}_{-0.11}$ & $^{+0.08}_{-0.07}$ & \\
 & 1.5 & 2.0 & 0.871 & $\pm$0.093 & $^{+0.087}_{-0.082}$ & $^{+0.035}_{-0.043}$ & \\
 & 2.0 & 2.4 & 0.74 & $\pm$0.11 & $^{+0.07}_{-0.15}$ & $^{+0.04}_{-0.03}$ & \\
\hline
$6<E_{T}^{\rm jet}<9$ & -1.5 & -1.0 & 0.546 & $\pm$0.066 & $^{+0.066}_{-0.111}$ & $^{+0.062}_{-0.046}$ & \\
 & -1.0 & -0.5 & 1.374 & $\pm$0.081 & $^{+0.166}_{-0.043}$ & $^{+0.097}_{-0.065}$ & \\
 & -0.5 & 0.0 & 2.067 & $\pm$0.097 & $^{+0.200}_{-0.163}$ & $^{+0.098}_{-0.091}$ & \\
 & 0.0 & 0.5 & 1.944 & $\pm$0.096 & $^{+0.217}_{-0.094}$ & $^{+0.065}_{-0.051}$ & \\
 & 0.5 & 1.0 & 1.429 & $\pm$0.093 & $^{+0.184}_{-0.059}$ & $^{+0.048}_{-0.040}$ & \\
 & 1.0 & 1.5 & 1.30 & $\pm$0.10 & $^{+0.12}_{-0.12}$ & $^{+0.05}_{-0.04}$ & \\
 & 1.5 & 2.0 & 0.542 & $\pm$0.075 & $^{+0.071}_{-0.080}$ & $^{+0.009}_{-0.027}$ & \\
 & 2.0 & 2.4 & 0.463 & $\pm$0.089 & $^{+0.051}_{-0.131}$ & $^{+0.023}_{-0.011}$ & \\
\hline
$E_{T}^{\rm jet}>9$ & -1.50 & -1.00 & & & & & \\
 & -1.0 & -0.5 & 0.377 & $\pm$0.051 & $^{+0.043}_{-0.054}$ & $^{+0.040}_{-0.039}$ & \\
 & -0.5 & 0.0 & 0.727 & $\pm$0.070 & $^{+0.076}_{-0.049}$ & $^{+0.067}_{-0.054}$ & \\
 & 0.0 & 0.5 & 0.950 & $\pm$0.074 & $^{+0.130}_{-0.045}$ & $^{+0.072}_{-0.070}$ & \\
 & 0.5 & 1.0 & 0.872 & $\pm$0.080 & $^{+0.132}_{-0.066}$ & $^{+0.065}_{-0.050}$ & \\
 & 1.0 & 1.5 & 0.559 & $\pm$0.079 & $^{+0.087}_{-0.071}$ & $^{+0.032}_{-0.028}$ & \\
 & 1.5 & 2.0 & 0.326 & $\pm$0.056 & $^{+0.064}_{-0.043}$ & $^{+0.027}_{-0.016}$ & \\
 & 2.0 & 2.4 & 0.276 & $\pm$0.063 & $^{+0.049}_{-0.066}$ & $^{+0.013}_{-0.019}$ & \\
\hline
    \end{tabular}
    \caption{The cross section $d\sigma / d \eta^{\rm jet}$ for events containing 
             at least one $D^*$ meson in different regions of $E_T^{\rm jet}$. 
             The statistical (stat.), systematic (syst.) and energy scale (E-scale) 
             uncertainties are shown separately.}
    \label{tbl:inclusive_eta}
  \end{center}
\end{table}
\begin{table}[htbp]
  \begin{center}
    \begin{tabular}{lr|cccc|cccc}
\hline
\multicolumn{2}{c|}{$E_{T}^{\rm jet}$ range} & \multicolumn{8}{c}{$d\sigma/dE_{T}^{\rm jet}$ $\pm$ stat. $\pm$ syst. $\pm$ E-scale (nb/GeV)} \\
\cline{3-10}
\multicolumn{2}{c|}{}&\multicolumn{4}{c|}{$D^{*}$-tagged jet} & \multicolumn{4}{c}{untagged jets} \\
\hline
6.00 & 9.00 &  0.889 & $\pm$0.028 & $^{+0.135}_{-0.039}$ & $^{+0.039}_{-0.028}$  & 0.704 & $\pm$0.028 & $^{+0.030}_{-0.045}$ & $^{+0.035}_{-0.035}$  \\
9.00 & 13.00 &  0.208 & $\pm$0.012 & $^{+0.033}_{-0.013}$ & $^{+0.015}_{-0.013}$  & 0.161 & $\pm$0.014 & $^{+0.006}_{-0.006}$ & $^{+0.010}_{-0.010}$  \\
13.00 & 18.00 &  0.0374 & $\pm$0.0059 & $^{+0.0064}_{-0.0054}$ & $^{+0.0049}_{-0.0035}$  & 0.0423 & $\pm$0.0076 & $^{+0.0122}_{-0.0048}$ & $^{+0.0049}_{-0.0043}$  \\
18.00 & 25.00 &  0.0145 & $\pm$0.0034 & $^{+0.0046}_{-0.0043}$ & $^{+0.0027}_{-0.0022}$  & 0.0124 & $\pm$0.0030 & $^{+0.0040}_{-0.0035}$ & $^{+0.0007}_{-0.0003}$  \\
\hline
    \end{tabular}
    \caption{The cross section $d\sigma / d E_{T}^{\rm jet}$ for $D^*$-tagged jets 
             and untagged jets for $-1.5 <\eta^{\rm jet} < 2.4$. The statistical 
             (stat.), systematic (syst.) and energy scale (E-scale) uncertainties 
	     are shown separately.}
    \label{tbl:inclusive_et_dstarother}
  \end{center}
\end{table}
\begin{table}[htbp]
  \begin{center}
    \begin{tabular}{lr|cccc|cccc}
\hline
\multicolumn{2}{c|}{$\eta^{\rm jet}$ range} & \multicolumn{8}{c}{$d\sigma/d\eta^{\rm jet}$ $\pm$ stat. $\pm$ syst. $\pm$ E-scale (nb)} \\
\cline{3-10}
\multicolumn{2}{c|}{}&\multicolumn{4}{c|}{$D^{*}$ matched jet} & \multicolumn{4}{c}{other jets} \\
\hline
-1.50 & -1.00 &  0.412 & $\pm$0.055 & $^{+0.060}_{-0.054}$ & $^{+0.043}_{-0.034}$  & 0.173 & $\pm$0.043 & $^{+0.058}_{-0.097}$ & $^{+0.022}_{-0.015}$  \\
-1.00 & -0.50 &  1.170 & $\pm$0.075 & $^{+0.236}_{-0.064}$ & $^{+0.089}_{-0.048}$  & 0.560 & $\pm$0.075 & $^{+0.039}_{-0.029}$ & $^{+0.035}_{-0.036}$  \\
-0.50 & 0.00 &  1.831 & $\pm$0.097 & $^{+0.324}_{-0.123}$ & $^{+0.084}_{-0.067}$  & 1.010 & $\pm$0.094 & $^{+0.081}_{-0.224}$ & $^{+0.053}_{-0.063}$  \\
0.00 & 0.50 &  1.78 & $\pm$0.11 & $^{+0.34}_{-0.12}$ & $^{+0.05}_{-0.04}$  & 1.113 & $\pm$0.095 & $^{+0.099}_{-0.056}$ & $^{+0.044}_{-0.044}$  \\
0.50 & 1.00 &  1.18 & $\pm$0.11 & $^{+0.30}_{-0.08}$ & $^{+0.03}_{-0.03}$  & 1.08 & $\pm$0.10 & $^{+0.10}_{-0.07}$ & $^{+0.05}_{-0.04}$  \\
1.00 & 1.50 &  0.99 & $\pm$0.13 & $^{+0.24}_{-0.11}$ & $^{+0.02}_{-0.01}$  & 1.014 & $\pm$0.098 & $^{+0.085}_{-0.129}$ & $^{+0.054}_{-0.046}$  \\
1.50 & 2.00 &  0.19 & $\pm$0.17 & $^{+0.22}_{-0.17}$ & $^{+0.00}_{-0.02}$  & 0.75 & $\pm$0.10 & $^{+0.10}_{-0.11}$ & $^{+0.01}_{-0.04}$  \\
2.00 & 2.40 &   & & &  & 0.70 & $\pm$0.13 & $^{+0.08}_{-0.20}$ & $^{+0.04}_{-0.02}$  \\
\hline
-1.50 & -1.00 &  0.386 & $\pm$0.052 & $^{+0.051}_{-0.051}$ & $^{+0.040}_{-0.032}$  & 0.160 & $\pm$0.040 & $^{+0.055}_{-0.089}$ & $^{+0.021}_{-0.014}$  \\
-1.00 & -0.50 &  0.935 & $\pm$0.060 & $^{+0.157}_{-0.037}$ & $^{+0.071}_{-0.038}$  & 0.438 & $\pm$0.058 & $^{+0.039}_{-0.023}$ & $^{+0.027}_{-0.028}$  \\
-0.50 & 0.00 &  1.367 & $\pm$0.073 & $^{+0.186}_{-0.060}$ & $^{+0.063}_{-0.050}$  & 0.714 & $\pm$0.067 & $^{+0.055}_{-0.148}$ & $^{+0.037}_{-0.045}$  \\
0.00 & 0.50 &  1.224 & $\pm$0.074 & $^{+0.182}_{-0.075}$ & $^{+0.037}_{-0.024}$  & 0.725 & $\pm$0.062 & $^{+0.065}_{-0.036}$ & $^{+0.029}_{-0.028}$  \\
0.50 & 1.00 &  0.771 & $\pm$0.069 & $^{+0.162}_{-0.050}$ & $^{+0.019}_{-0.017}$  & 0.665 & $\pm$0.063 & $^{+0.055}_{-0.039}$ & $^{+0.033}_{-0.025}$  \\
1.00 & 1.50 &  0.651 & $\pm$0.086 & $^{+0.132}_{-0.076}$ & $^{+0.015}_{-0.010}$  & 0.623 & $\pm$0.061 & $^{+0.049}_{-0.080}$ & $^{+0.033}_{-0.028}$  \\
1.50 & 2.00 &  0.113 & $\pm$0.100 & $^{+0.154}_{-0.098}$ & $^{+0.001}_{-0.013}$  & 0.473 & $\pm$0.066 & $^{+0.064}_{-0.070}$ & $^{+0.008}_{-0.022}$  \\
2.00 & 2.40 &   & & &  & 0.466 & $\pm$0.089 & $^{+0.049}_{-0.131}$ & $^{+0.023}_{-0.011}$  \\
\hline
-1.50 & -1.00 &   & & &  &  & & &  \\
-1.00 & -0.50 &  0.239 & $\pm$0.036 & $^{+0.039}_{-0.023}$ & $^{+0.023}_{-0.023}$  & 0.148 & $\pm$0.043 & $^{+0.021}_{-0.091}$ & $^{+0.021}_{-0.019}$  \\
-0.50 & 0.00 &  0.514 & $\pm$0.050 & $^{+0.064}_{-0.047}$ & $^{+0.047}_{-0.038}$  & 0.194 & $\pm$0.055 & $^{+0.058}_{-0.015}$ & $^{+0.018}_{-0.015}$  \\
0.00 & 0.50 &  0.620 & $\pm$0.055 & $^{+0.096}_{-0.045}$ & $^{+0.051}_{-0.048}$  & 0.328 & $\pm$0.051 & $^{+0.053}_{-0.018}$ & $^{+0.020}_{-0.021}$  \\
0.50 & 1.00 &  0.520 & $\pm$0.056 & $^{+0.118}_{-0.042}$ & $^{+0.041}_{-0.026}$  & 0.358 & $\pm$0.063 & $^{+0.038}_{-0.052}$ & $^{+0.023}_{-0.026}$  \\
1.00 & 1.50 &  0.231 & $\pm$0.057 & $^{+0.044}_{-0.041}$ & $^{+0.015}_{-0.011}$  & 0.335 & $\pm$0.055 & $^{+0.067}_{-0.039}$ & $^{+0.016}_{-0.018}$  \\
1.50 & 2.00 &  0.140 & $\pm$0.080 & $^{+0.080}_{-0.078}$ & $^{+0.022}_{-0.012}$  & 0.263 & $\pm$0.048 & $^{+0.048}_{-0.033}$ & $^{+0.021}_{-0.013}$  \\
2.00 & 2.40 &   & & &  & 0.275 & $\pm$0.062 & $^{+0.049}_{-0.065}$ & $^{+0.013}_{-0.019}$  \\
\hline
    \end{tabular}
    \caption{The cross section $d\sigma / d \eta^{\rm jet}$ for $D^*$-tagged jets 
             and untagged jets for $E_T^{\rm jet}>6$\,GeV. The statistical 
             (stat.), systematic (syst.) and energy scale (E-scale) uncertainties 
	     are shown separately.}
    \label{tbl:inclusive_eta_dstarother}
  \end{center}
\end{table}
\begin{table}[htbp]
  \begin{center}
    \begin{tabular}{lr|lr|ccccc}
\hline
\multicolumn{2}{c|}{$p_{T}(D^{*})$ range} & \multicolumn{2}{c|}{$E_{T}^{\rm jet}$ range} & $d\sigma/dE_{T}^{\rm jet}$ & $\pm$ stat. & $\pm$ syst. & $\pm$ E-scale &
(nb/GeV) \\
\hline
3.25 & 5.00 & 6.0 & 9.0 & 0.907 & $\pm$0.032 & $^{+0.130}_{-0.042}$ & $^{+0.083}_{-0.067}$ & \\
\multicolumn{2}{c|}{} & 9.0 & 13.0 & 0.137 & $\pm$0.015 & $^{+0.018}_{-0.012}$ & $^{+0.014}_{-0.013}$ & \\
\multicolumn{2}{c|}{} & 13.0 & 18.0 & 0.0279 & $\pm$0.0085 & $^{+0.0134}_{-0.0055}$ & $^{+0.0027}_{-0.0040}$ & \\
\multicolumn{2}{c|}{} & 18.0 & 25.0 & 0.0095 & $\pm$0.0047 & $^{+0.0050}_{-0.0051}$ & $^{+-0.0006}_{-0.0018}$ & \\
\hline
5.00 & 8.00 & 6.0 & 9.0 & 0.510 & $\pm$0.016 & $^{+0.043}_{-0.039}$ & $^{+0.004}_{-0.000}$ & \\
\multicolumn{2}{c|}{} & 9.0 & 13.0 & 0.1432 & $\pm$0.0091 & $^{+0.0127}_{-0.0050}$ & $^{+0.0142}_{-0.0112}$ & \\
\multicolumn{2}{c|}{} & 13.0 & 18.0 & 0.0200 & $\pm$0.0040 & $^{+0.0060}_{-0.0021}$ & $^{+0.0030}_{-0.0017}$ & \\
\multicolumn{2}{c|}{} & 18.0 & 25.0 & 0.0093 & $\pm$0.0034 & $^{+0.0070}_{-0.0038}$ & $^{+0.0024}_{-0.0025}$ & \\
\hline
8.00 & 20.00 & 6.0 & 9.0 & 0.0377 & $\pm$0.0036 & $^{+0.0021}_{-0.0071}$ & $^{+-0.0018}_{0.0014}$ & \\
\multicolumn{2}{c|}{} & 9.0 & 13.0 & 0.0731 & $\pm$0.0050 & $^{+0.0016}_{-0.0087}$ & $^{+-0.0009}_{0.0005}$ & \\
\multicolumn{2}{c|}{} & 13.0 & 18.0 & 0.0338 & $\pm$0.0036 & $^{+0.0021}_{-0.0017}$ & $^{+0.0039}_{-0.0031}$ & \\
\multicolumn{2}{c|}{} & 18.0 & 25.0 & 0.0092 & $\pm$0.0019 & $^{+0.0012}_{-0.0012}$ & $^{+0.0012}_{-0.0004}$ & \\
\hline
    \end{tabular}
    \caption{The cross section $d\sigma / d E_{T}^{\rm jet}$ in bins of $p_{T}^{D^*}$
             The statistical (stat.), systematic (syst.) and energy scale (E-scale) 
	     uncertainties are shown separately.}
    \label{tbl:inclusive_et_ptds}
  \end{center}
\end{table}
\begin{table}[htbp]
  \begin{center}
    \begin{tabular}{lr|lr|ccccc}
\hline
\multicolumn{2}{c|}{$p_{T}(D^{*})$ range} & \multicolumn{2}{c|}{$\eta^{\rm jet}$ range} & $d\sigma/d\eta^{\rm jet}$ & $\pm$ stat. & $\pm$ syst. & $\pm$ E-scale & (nb) \\
\hline
3.25 & 5.00 & -1.5 & -1.0 & 0.360 & $\pm$0.058 & $^{+0.074}_{-0.117}$ & $^{+0.071}_{-0.051}$ & \\
 & & -1.0 & -0.5 & 0.903 & $\pm$0.075 & $^{+0.282}_{-0.105}$ & $^{+0.113}_{-0.083}$ & \\
 & & -0.5 & 0.0 & 1.465 & $\pm$0.096 & $^{+0.162}_{-0.184}$ & $^{+0.140}_{-0.122}$ & \\
 & & 0.0 & 0.5 & 1.453 & $\pm$0.094 & $^{+0.207}_{-0.167}$ & $^{+0.110}_{-0.103}$ & \\
 & & 0.5 & 1.0 & 1.063 & $\pm$0.097 & $^{+0.244}_{-0.046}$ & $^{+0.094}_{-0.074}$ & \\
 & & 1.0 & 1.5 & 0.98 & $\pm$0.11 & $^{+0.16}_{-0.20}$ & $^{+0.08}_{-0.06}$ & \\
 & & 1.5 & 2.0 & 0.391 & $\pm$0.076 & $^{+0.078}_{-0.048}$ & $^{+0.021}_{-0.025}$ & \\
 & & 2.0 & 2.4 & 0.335 & $\pm$0.092 & $^{+0.102}_{-0.129}$ & $^{+0.018}_{-0.013}$ & \\
\hline
5.00 & 8.00 & -1.5 & -1.0 & 0.211 & $\pm$0.033 & $^{+0.081}_{-0.051}$ & $^{+0.014}_{-0.007}$ & \\
 & & -1.0 & -0.5 & 0.602 & $\pm$0.046 & $^{+0.050}_{-0.030}$ & $^{+0.029}_{-0.018}$ & \\
 & & -0.5 & 0.0 & 0.981 & $\pm$0.056 & $^{+0.066}_{-0.048}$ & $^{+0.030}_{-0.024}$ & \\
 & & 0.0 & 0.5 & 1.015 & $\pm$0.057 & $^{+0.105}_{-0.074}$ & $^{+0.026}_{-0.022}$ & \\
 & & 0.5 & 1.0 & 0.745 & $\pm$0.051 & $^{+0.107}_{-0.037}$ & $^{+0.022}_{-0.016}$ & \\
 & & 1.0 & 1.5 & 0.534 & $\pm$0.053 & $^{+0.052}_{-0.050}$ & $^{+0.018}_{-0.014}$ & \\
 & & 1.5 & 2.0 & 0.231 & $\pm$0.033 & $^{+0.033}_{-0.018}$ & $^{+0.011}_{-0.009}$ & \\
 & & 2.0 & 2.4 & 0.240 & $\pm$0.045 & $^{+0.035}_{-0.049}$ & $^{+0.011}_{-0.009}$ & \\
\hline
8.0 & 20.0 & -1.50 & -1.00 & & & & & \\
 & & -1.0 & -0.5 & 0.101 & $\pm$0.020 & $^{+0.009}_{-0.021}$ & $^{+0.001}_{-0.002}$ & \\
 & & -0.5 & 0.0 & 0.217 & $\pm$0.024 & $^{+0.010}_{-0.017}$ & $^{+0.002}_{-0.002}$ & \\
 & & 0.0 & 0.5 & 0.274 & $\pm$0.027 & $^{+0.019}_{-0.011}$ & $^{+0.002}_{-0.001}$ & \\
 & & 0.5 & 1.0 & 0.296 & $\pm$0.028 & $^{+0.015}_{-0.036}$ & $^{+0.003}_{-0.001}$ & \\
 & & 1.0 & 1.5 & 0.178 & $\pm$0.024 & $^{+0.008}_{-0.018}$ & $^{+0.003}_{-0.003}$ & \\
 & & 1.5 & 2.0 & 0.108 & $\pm$0.021 & $^{+0.013}_{-0.020}$ & $^{+0.000}_{-0.004}$ & \\
 & & 2.0 & 2.4 & 0.092 & $\pm$0.020 & $^{+0.014}_{-0.019}$ & $^{+0.004}_{-0.005}$ & \\
\hline
    \end{tabular}
    \caption{The cross section $d\sigma / d \eta^{\rm jet}$ in bins of $p_{T}^{D^*}$. 
             The statistical (stat.), systematic (syst.) and energy scale (E-scale) 
	     uncertainties are shown separately.}
    \label{tbl:inclusive_eta_ptds}
  \end{center}
\end{table}
\begin{table}[htbp]
  \begin{center}
    \begin{tabular}{lr|ccccc}
\hline
\multicolumn{2}{c|}{$x_{\gamma}(D^{*},{\rm jet})$ range} & $d\sigma/dx_{\gamma}(D^{*},{\rm jet})$ & $\pm$ stat. & $\pm$ syst. & $\pm$ E-scale & (nb) \\
\hline
0.000 & 0.250 & 1.23 & $\pm$0.24 & $^{+0.13}_{-0.36}$ & $^{+0.03}_{-0.06}$ & \\
0.250 & 0.375 & 2.15 & $\pm$0.28 & $^{+0.37}_{-0.30}$ & $^{+0.09}_{-0.11}$ & \\
0.375 & 0.500 & 2.85 & $\pm$0.31 & $^{+0.36}_{-0.25}$ & $^{+0.09}_{-0.10}$ & \\
0.500 & 0.625 & 3.82 & $\pm$0.30 & $^{+0.46}_{-0.36}$ & $^{+0.04}_{-0.06}$ & \\
0.625 & 0.750 & 5.71 & $\pm$0.32 & $^{+0.23}_{-0.57}$ & $^{+0.01}_{-0.09}$ & \\
0.750 & 1.000 & 2.82 & $\pm$0.15 & $^{+0.25}_{-0.25}$ & $^{+0.37}_{-0.30}$ & \\
\hline
    \end{tabular}
    \caption{The cross section $d\sigma / d x_{\gamma}^{\rm obs}(D^*,other~jet)$. The 
             statistical (stat.), systematic (syst.) and energy scale (E-scale) 
	     uncertainties are shown separately.}
    \label{tbl:inclusive_xgamma_dsjet}
  \end{center}
\end{table}
\begin{table}[htbp]
  \begin{center}
    \begin{tabular}{lr|ccccc}
\hline
\multicolumn{2}{c|}{$x_{\gamma}^{\rm obs}$ range} & $d\sigma/dx_{\gamma}^{\rm obs}$ & $\pm$ stat. & $\pm$ syst. & $\pm$ E-scale & (nb) \\
\hline
0.000 & 0.250 & 0.19 & $\pm$0.15 & $^{+0.34}_{-0.07}$ & $^{+0.02}_{-0.01}$ & \\
0.250 & 0.375 & 0.78 & $\pm$0.22 & $^{+0.26}_{-0.23}$ & $^{+0.08}_{-0.08}$ & \\
0.375 & 0.500 & 0.99 & $\pm$0.19 & $^{+0.24}_{-0.23}$ & $^{+0.08}_{-0.09}$ & \\
0.500 & 0.625 & 1.52 & $\pm$0.22 & $^{+0.14}_{-0.33}$ & $^{+0.12}_{-0.12}$ & \\
0.625 & 0.750 & 2.03 & $\pm$0.24 & $^{+0.21}_{-0.23}$ & $^{+0.18}_{-0.16}$ & \\
0.750 & 1.000 & 4.35 & $\pm$0.21 & $^{+0.41}_{-0.16}$ & $^{+0.28}_{-0.26}$ & \\
\hline
    \end{tabular}
    \caption{The dijet cross section $d\sigma / d x_{\gamma}^{\rm obs}$, for events 
             containing at least one $D^*$ meson. The statistical (stat.), systematic 
	     (syst.) and energy scale (E-scale) uncertainties are shown separately.}
    \label{tbl:inclusive_xgamma}
  \end{center}
\end{table}
\begin{table}[htbp]
  \begin{center}
    \begin{tabular}{lr|ccccc}
\hline
\multicolumn{2}{c|}{$\Delta\phi^{\rm jj}$ range} & $d\sigma/d\Delta\phi^{\rm jj}$ & $\pm$ stat. & $\pm$ syst. & $\pm$ E-scale & (nb/rad.) \\
\hline
\multicolumn{2}{c}{}& \multicolumn{5}{c}{$0<x_{\gamma}^{\rm obs}<1$} \\
\hline
0.00 & 1.57 &  0.020 & $\pm$0.011 & $^{+0.012}_{-0.012}$ & $^{+0.001}_{-0.003}$ &  \\
1.57 & 2.09 &  0.149 & $\pm$0.035 & $^{+0.051}_{-0.046}$ & $^{+0.025}_{-0.014}$ &  \\
2.09 & 2.36 &  0.318 & $\pm$0.074 & $^{+0.057}_{-0.092}$ & $^{+0.028}_{-0.029}$ &  \\
2.36 & 2.62 &  0.67 & $\pm$0.10 & $^{+0.17}_{-0.10}$ & $^{+0.07}_{-0.05}$ &  \\
2.62 & 2.88 &  1.50 & $\pm$0.14 & $^{+0.18}_{-0.12}$ & $^{+0.11}_{-0.11}$ &  \\
2.88 & 3.14 &  4.06 & $\pm$0.22 & $^{+0.39}_{-0.32}$ & $^{+0.24}_{-0.23}$ &  \\
\hline
\multicolumn{2}{c}{}& \multicolumn{5}{c}{$x_{\gamma}^{\rm obs}>0.75$} \\
\hline
0.00 & 1.57 &  0.0158 & $\pm$0.0068 & $^{+0.0058}_{-0.0115}$ & $^{+0.0029}_{-0.0017}$ &  \\
1.57 & 2.09 &  0.036 & $\pm$0.014 & $^{+0.006}_{-0.028}$ & $^{+0.007}_{-0.003}$ &  \\
2.09 & 2.36 &  0.126 & $\pm$0.038 & $^{+0.032}_{-0.027}$ & $^{+0.011}_{-0.012}$ &  \\
2.36 & 2.62 &  0.389 & $\pm$0.055 & $^{+0.077}_{-0.021}$ & $^{+0.036}_{-0.028}$ &  \\
2.62 & 2.88 &  0.829 & $\pm$0.091 & $^{+0.105}_{-0.066}$ & $^{+0.055}_{-0.055}$ &  \\
2.88 & 3.14 &  2.68 & $\pm$0.16 & $^{+0.28}_{-0.13}$ & $^{+0.14}_{-0.13}$ &  \\
\hline
\multicolumn{2}{c}{}& \multicolumn{5}{c}{$x_{\gamma}^{\rm obs}<0.75$} \\
\hline
0.00 & 1.57 &  0.0066 & $\pm$0.0093 & $^{+0.0141}_{-0.0074}$ & $^{+0.0004}_{-0.0010}$ &  \\
1.57 & 2.09 &  0.115 & $\pm$0.034 & $^{+0.049}_{-0.029}$ & $^{+0.017}_{-0.012}$ &  \\
2.09 & 2.36 &  0.190 & $\pm$0.062 & $^{+0.042}_{-0.082}$ & $^{+0.018}_{-0.018}$ &  \\
2.36 & 2.62 &  0.280 & $\pm$0.083 & $^{+0.157}_{-0.104}$ & $^{+0.030}_{-0.020}$ &  \\
2.62 & 2.88 &  0.67 & $\pm$0.11 & $^{+0.10}_{-0.10}$ & $^{+0.06}_{-0.06}$ &  \\
2.88 & 3.14 &  1.37 & $\pm$0.15 & $^{+0.14}_{-0.26}$ & $^{+0.11}_{-0.10}$ &  \\
\hline
    \end{tabular}
    \caption{The dijet cross section $d\sigma / d \phi^{\rm jj}$, for events 
             containing at least one $D^*$ meson for all $x_{\gamma}^{\rm obs}$, 
	     and for direct-enriched ($x_{\gamma}^{\rm obs}>0.75$) and 
	     resolved-enriched ($x_{\gamma}^{\rm obs}<0.75$) samples. The 
	     statistical (stat.), systematic (syst.) and energy scale (E-scale) 
	     uncertainties are shown separately.}
    \label{tbl:inclusive_dphi}
  \end{center}
\end{table}
\begin{table}[htbp]
  \begin{center}
    \begin{tabular}{lr|ccccc}
\hline
\multicolumn{2}{c|}{$p_{T}^{\rm jj}$ range} & $d\sigma/dp_{T}^{\rm jj}$ & $\pm$ stat. & $\pm$ syst. & $\pm$ E-scale & (nb/GeV) \\
\hline
\multicolumn{2}{c}{}& \multicolumn{5}{c}{$0<x_{\gamma}^{\rm obs}<1$} \\
\hline
 0 &  3 &  0.282 & $\pm$0.018 & $^{+0.027}_{-0.021}$ & $^{+0.021}_{-0.019}$ &  \\
 3 &  6 &  0.188 & $\pm$0.014 & $^{+0.033}_{-0.016}$ & $^{+0.013}_{-0.013}$ &  \\
 6 &  9 &  0.0762 & $\pm$0.0095 & $^{+0.0111}_{-0.0096}$ & $^{+0.0048}_{-0.0049}$ &  \\
 9 & 12 &  0.0351 & $\pm$0.0079 & $^{+0.0070}_{-0.0127}$ & $^{+0.0026}_{-0.0025}$ &  \\
12 & 18 &  0.0109 & $\pm$0.0036 & $^{+0.0037}_{-0.0046}$ & $^{+0.0006}_{-0.0001}$ &  \\
\hline
\multicolumn{2}{c}{}& \multicolumn{5}{c}{$x_{\gamma}^{\rm obs}>0.75$} \\
\hline
 0 &  3 &  0.203 & $\pm$0.013 & $^{+0.021}_{-0.018}$ & $^{+0.012}_{-0.012}$ &  \\
 3 &  6 &  0.1074 & $\pm$0.0092 & $^{+0.0136}_{-0.0090}$ & $^{+0.0061}_{-0.0058}$ &  \\
 6 &  9 &  0.0300 & $\pm$0.0049 & $^{+0.0060}_{-0.0053}$ & $^{+0.0020}_{-0.0018}$ &  \\
 9 & 12 &  0.0141 & $\pm$0.0045 & $^{+0.0033}_{-0.0057}$ & $^{+0.0019}_{-0.0008}$ &  \\
12 & 18 &  0.0066 & $\pm$0.0018 & $^{+0.0008}_{-0.0040}$ & $^{+0.0003}_{0.0003}$ &  \\
\hline
\multicolumn{2}{c}{}& \multicolumn{5}{c}{$x_{\gamma}^{\rm obs}<0.75$} \\
\hline
 0 &  3 &  0.078 & $\pm$0.011 & $^{+0.009}_{-0.014}$ & $^{+0.008}_{-0.006}$ &  \\
 3 &  6 &  0.082 & $\pm$0.011 & $^{+0.022}_{-0.011}$ & $^{+0.008}_{-0.007}$ &  \\
 6 &  9 &  0.0479 & $\pm$0.0087 & $^{+0.0099}_{-0.0106}$ & $^{+0.0030}_{-0.0033}$ &  \\
 9 & 12 &  0.0207 & $\pm$0.0063 & $^{+0.0053}_{-0.0098}$ & $^{+0.0008}_{-0.0016}$ &  \\
12 & 18 &  0.0046 & $\pm$0.0031 & $^{+0.0037}_{-0.0025}$ & $^{+0.0003}_{-0.0004}$ &  \\
\hline
    \end{tabular}
    \caption{The dijet cross section $d\sigma / d p_{T}^{\rm jj}$, for events 
             containing at least one $D^*$ meson for all $x_{\gamma}^{\rm obs}$, 
	     and for direct-enriched ($x_{\gamma}^{\rm obs}>0.75$) and 
	     resolved-enriched ($x_{\gamma}^{\rm obs}<0.75$) samples. The 
	     statistical (stat.), systematic (syst.) and energy scale (E-scale) 
	     uncertainties are shown separately.}
    \label{tbl:inclusive_ptjj}
  \end{center}
\end{table}
\begin{table}[htbp]
  \begin{center}
    \begin{tabular}{lr|ccccc}
\hline
\multicolumn{2}{c|}{$m^{\rm jj}$ range} & $d\sigma/dm^{\rm jj}$ & $\pm$ stat. & $\pm$ syst. & $\pm$ E-scale & (nb/GeV) \\
\hline
\multicolumn{2}{c}{}& \multicolumn{5}{c}{$0<x_{\gamma}^{\rm obs}<1$} \\
\hline
15 & 20 &  0.1299 & $\pm$0.0088 & $^{+0.0095}_{-0.0124}$ & $^{+0.0080}_{-0.0076}$ &  \\
20 & 25 &  0.0823 & $\pm$0.0076 & $^{+0.0082}_{-0.0076}$ & $^{+0.0060}_{-0.0059}$ &  \\
25 & 30 &  0.0484 & $\pm$0.0062 & $^{+0.0106}_{-0.0052}$ & $^{+0.0049}_{-0.0023}$ &  \\
30 & 35 &  0.0328 & $\pm$0.0056 & $^{+0.0048}_{-0.0080}$ & $^{+0.0025}_{-0.0043}$ &  \\
35 & 40 &  0.0223 & $\pm$0.0048 & $^{+0.0018}_{-0.0054}$ & $^{+0.0033}_{0.0009}$ &  \\
40 & 50 &  0.0081 & $\pm$0.0022 & $^{+0.0017}_{-0.0016}$ & $^{+0.0006}_{-0.0010}$ &  \\
\hline
\multicolumn{2}{c}{}& \multicolumn{5}{c}{$x_{\gamma}^{\rm obs}>0.75$} \\
\hline
15 & 20 &  0.0788 & $\pm$0.0056 & $^{+0.0064}_{-0.0065}$ & $^{+0.0033}_{-0.0035}$ &  \\
20 & 25 &  0.0470 & $\pm$0.0049 & $^{+0.0061}_{-0.0029}$ & $^{+0.0042}_{-0.0028}$ &  \\
25 & 30 &  0.0319 & $\pm$0.0045 & $^{+0.0049}_{-0.0024}$ & $^{+0.0026}_{-0.0015}$ &  \\
30 & 35 &  0.0221 & $\pm$0.0045 & $^{+0.0042}_{-0.0045}$ & $^{+0.0018}_{-0.0031}$ &  \\
35 & 40 &  0.0136 & $\pm$0.0035 & $^{+0.0019}_{-0.0045}$ & $^{+0.0022}_{0.0003}$ &  \\
40 & 50 &  0.0040 & $\pm$0.0016 & $^{+0.0021}_{-0.0013}$ & $^{+0.0003}_{-0.0004}$ &  \\
\hline
\multicolumn{2}{c}{}& \multicolumn{5}{c}{$x_{\gamma}^{\rm obs}<0.75$} \\
\hline
15 & 20 &  0.0512 & $\pm$0.0069 & $^{+0.0055}_{-0.0097}$ & $^{+0.0051}_{-0.0043}$ &  \\
20 & 25 &  0.0357 & $\pm$0.0060 & $^{+0.0043}_{-0.0083}$ & $^{+0.0016}_{-0.0034}$ &  \\
25 & 30 &  0.0166 & $\pm$0.0041 & $^{+0.0059}_{-0.0036}$ & $^{+0.0023}_{-0.0008}$ &  \\
30 & 35 &  0.0109 & $\pm$0.0033 & $^{+0.0030}_{-0.0043}$ & $^{+0.0007}_{-0.0013}$ &  \\
35 & 40 &  0.0086 & $\pm$0.0033 & $^{+0.0022}_{-0.0027}$ & $^{+0.0011}_{0.0006}$ &  \\
40 & 50 &  0.0038 & $\pm$0.0015 & $^{+0.0009}_{-0.0013}$ & $^{+0.0004}_{-0.0006}$ &  \\
\hline
    \end{tabular}
    \caption{The dijet cross section $d\sigma / d M^{\rm jj}$, for events 
             containing at least one $D^*$ meson for all $x_{\gamma}^{\rm obs}$, 
	     and for direct-enriched ($x_{\gamma}^{\rm obs}>0.75$) and 
	     resolved-enriched ($x_{\gamma}^{\rm obs}<0.75$) samples. The 
	     statistical (stat.), systematic (syst.) and energy scale (E-scale) 
	     uncertainties are shown separately.}
    \label{tbl:inclusive_mjj}
  \end{center}
\end{table}
}

\clearpage

{
\begin{figure}[p]
  \vfill
  \begin{center}
    \includegraphics[width=\linewidth]{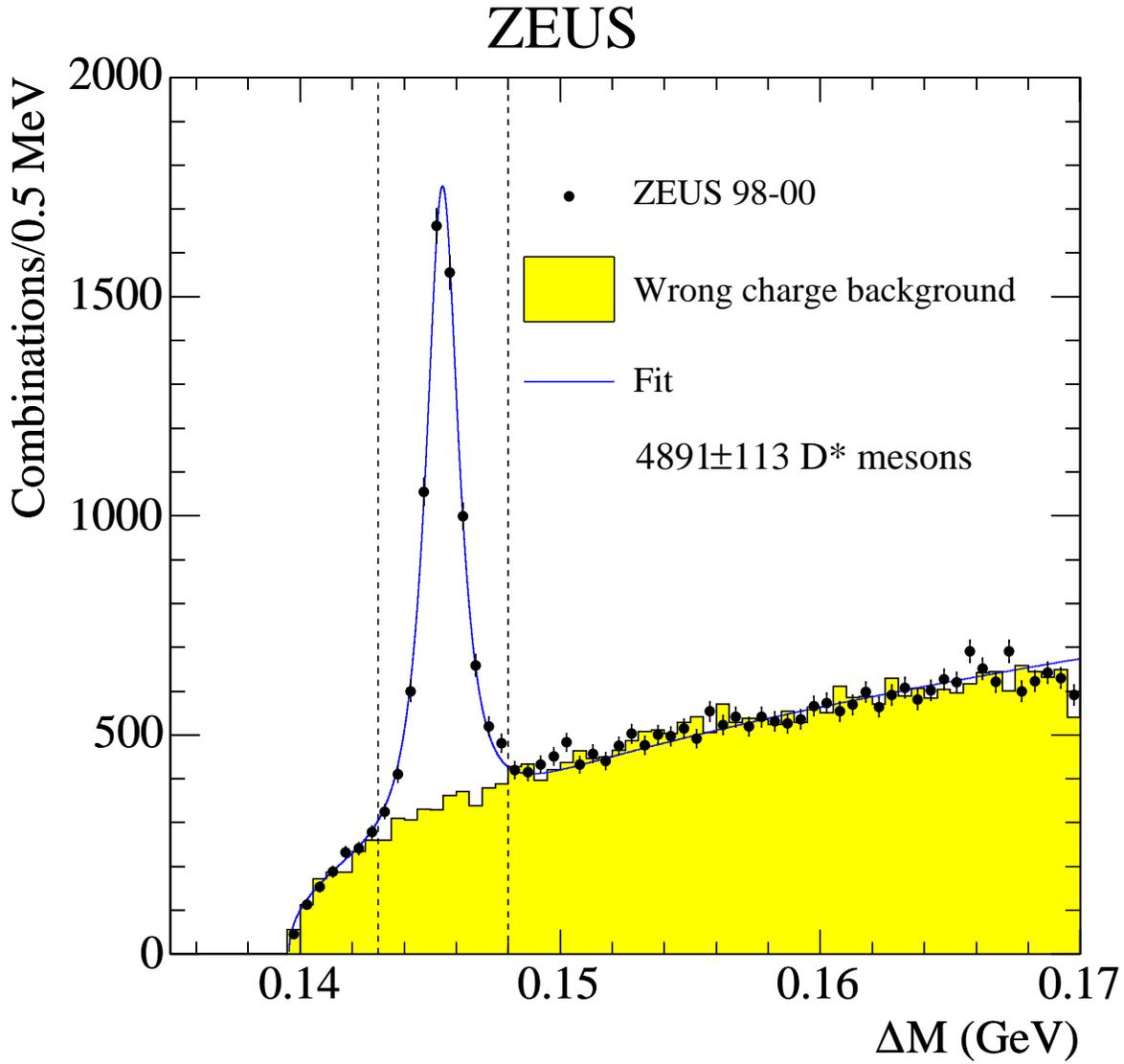}
  \end{center}
  \caption{
    The distribution of the mass difference, $\Delta M = M(K\pi\pi_{s})-M(K\pi)$, for 
    $D^*$ candidates with a single jet. The $D^{*\pm}$ candidates 
    (dots) are shown compared to the wrong charge combinations (histogram). The dashed 
    vertical lines show the signal region, 0.143 $< \Delta M <$ 0.148 GeV. The number of 
    $D^*$ mesons is determined by subtracting the wrong charge background as described in 
    Section~\ref{sec:dstar-rec}. The fit is for illustrative purposes only. 
  }
  \label{fig-deltam}
  \vfill
\end{figure}

\begin{figure}[p]
  \vfill
  \begin{center}
    \includegraphics[width=\linewidth]{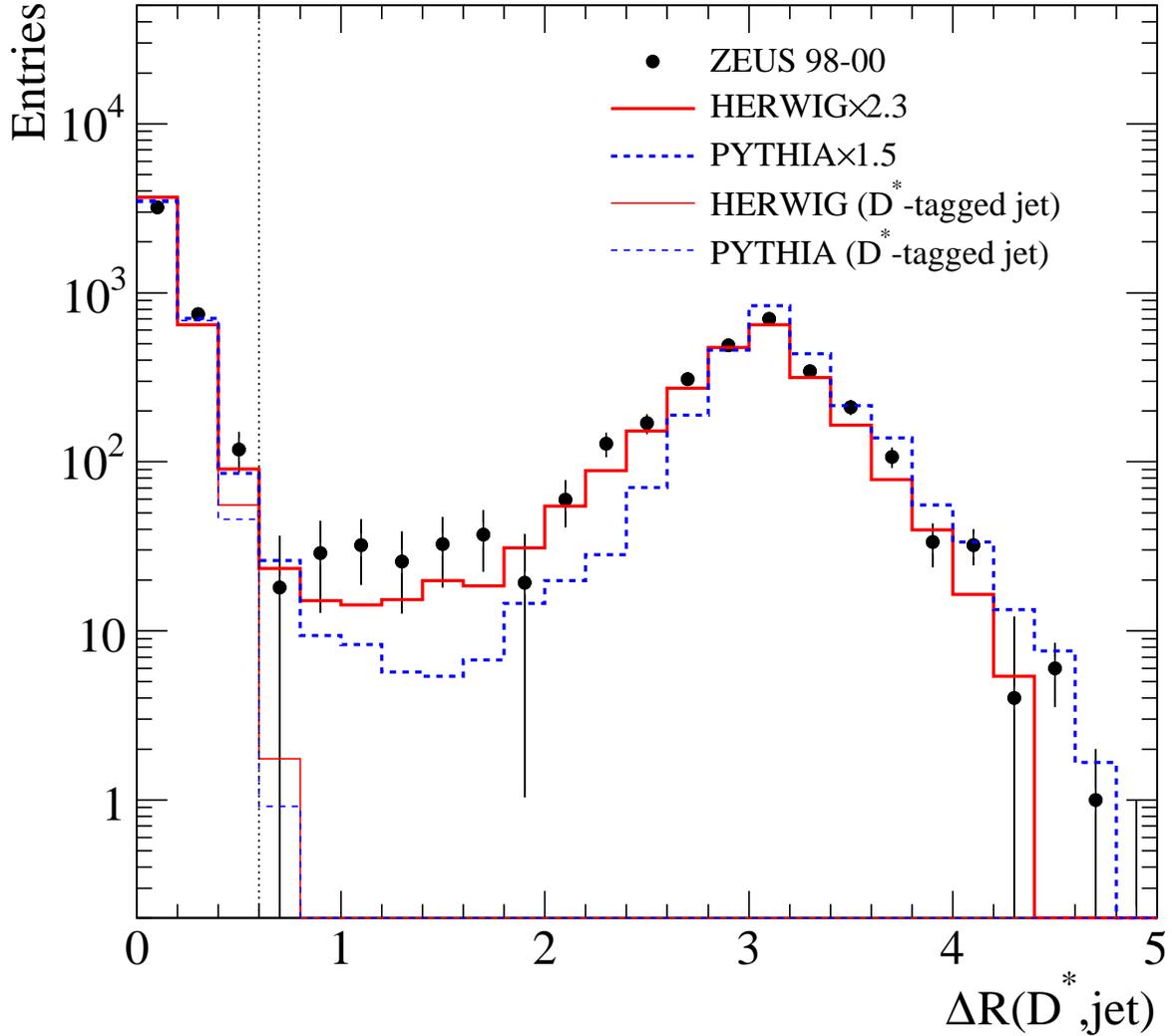}
  \end{center}
  \caption{
    The distribution of $\Delta R$ between $D^*$ mesons and each jet in the event. The 
    data (dots) are compared to {\sc Herwig} (solid line) and {\sc Pythia} (dashed line) 
    MC predictions. The MC predictions are area normalised to the data using the 
    normalisation factors shown in the figure. The dotted vertical line indicates the 
    $\Delta R = 0.6$ cut which separates $D^*$-tagged jets from untagged jets. The 
    $\Delta R(D^{*},{\rm jet})$ distribution for $D^*$-tagged jets for the MC hadron-level 
    predictions is shown for {\sc Herwig} (thin solid line) and {\sc Pythia} (thin dashed 
    line). The hadron-level predictions are area normalised to the data in the region 
    $\Delta R(D^{*},{\rm jet})<0.6$
  }
  \label{fig-deltar}
  \vfill
\end{figure}

\begin{figure}[p]
  \vfill
  \begin{center}
    \includegraphics[width=13cm]{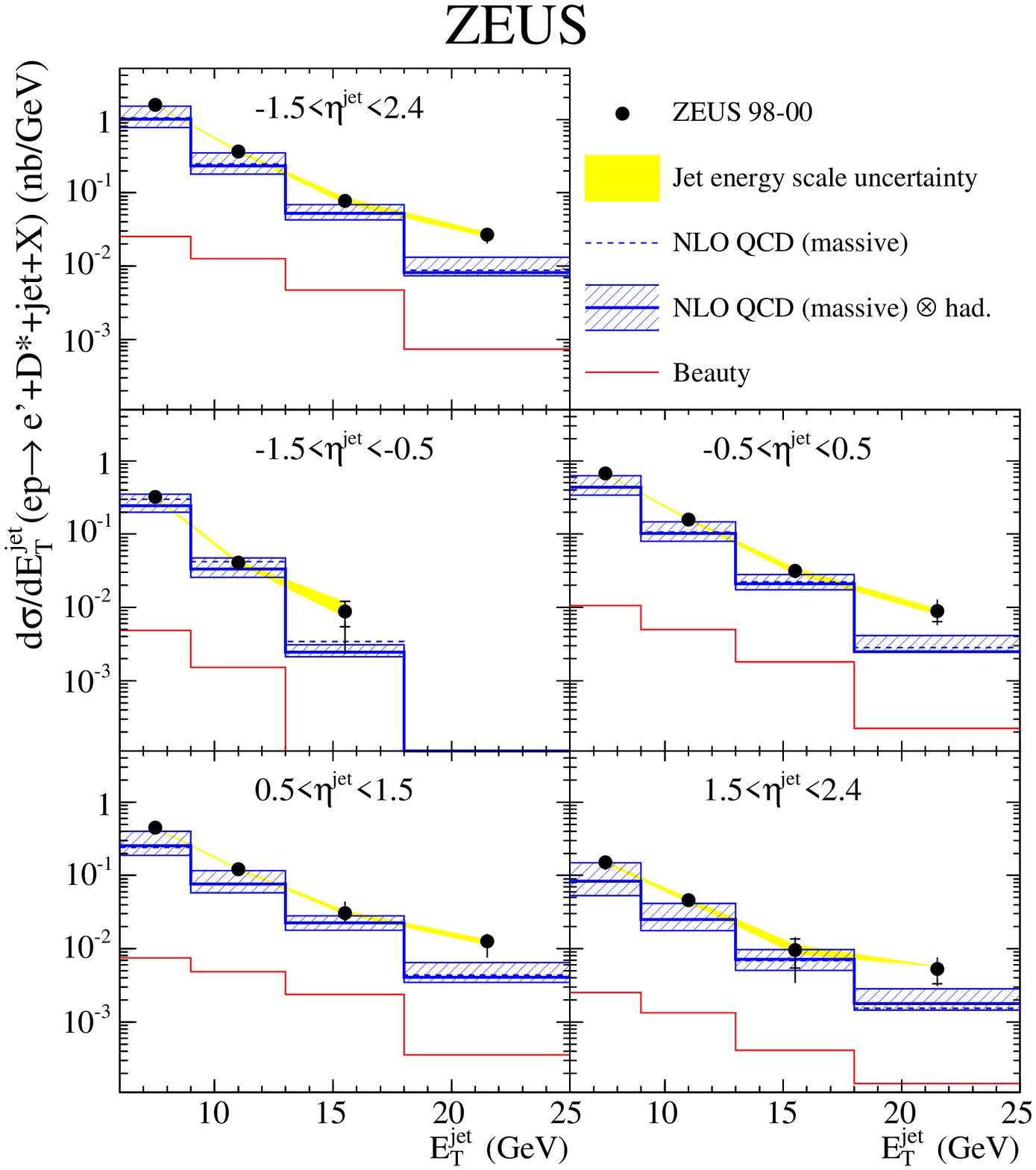}
  \end{center}
  \caption{
Cross-section $d\sigma / d E_{T}^{\rm jet}$ for jets in events (dots) containing 
at least one $D^*$ meson for different regions in $\eta^{\rm jet}$. The comparison 
is made to massive QCD predictions with (solid line) and without (dashed line) 
hadronisation corrections applied. The theoretical uncertainties (hatched band) 
come from the change in scales simultaneously with the change 
in charm mass. The beauty component is also shown (lower histogram).
  }
  \label{fig-et}
  \vfill
\end{figure}

\begin{figure}[p]
  \vfill
  \begin{center}
    \includegraphics[width=\linewidth]{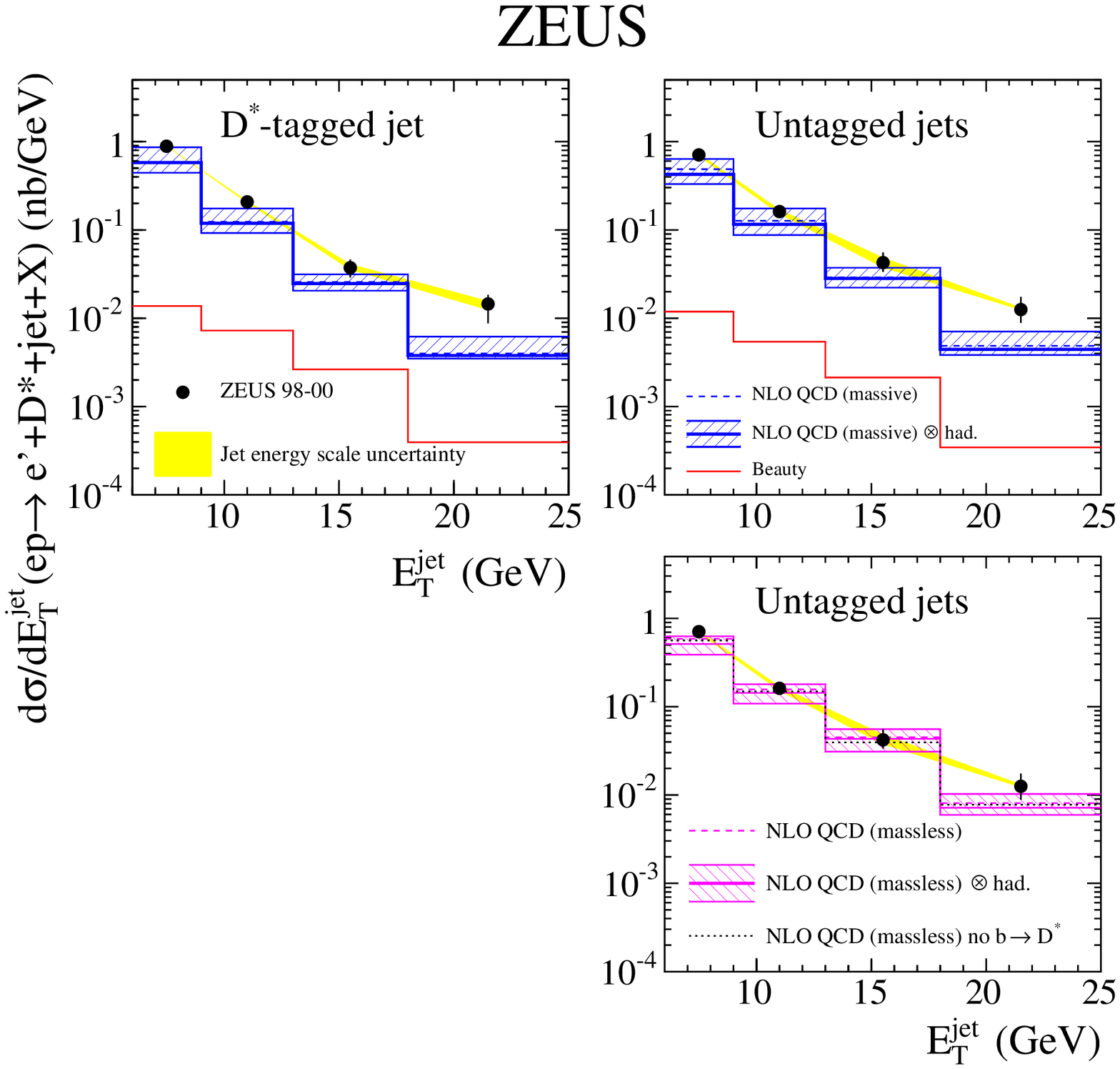}
  \end{center}
  \caption{
Cross-section $d\sigma / d E_{T}^{\rm jet}$ for $D^*$-tagged jets and untagged jets 
(dots). The comparison is made to massive QCD predictions with (solid line) and without 
(dotted line) hadronisation corrections applied. The beauty component is also shown (lower histogram). 
For the untagged jet distribution, the massless QCD predictions are also shown with (solid line) 
and without (dashed line) hadronisation corrections applied. 
The theoretical uncertainties (hatched bands) come, in the case of the massive calculations, from 
changing renormalisation and factorisation scales as well as the charm mass simultaneously.
In the case of the massless calculations, they come from changing the
scales only.
The prediction with no component from $b$-quark fragmentation to a $D^*$ is also shown (dotted line).
  }
  \label{fig-et_dstarother}
  \vfill
\end{figure}

\begin{figure}[p]
  \vfill
  \begin{center}
    \includegraphics[width=\linewidth]{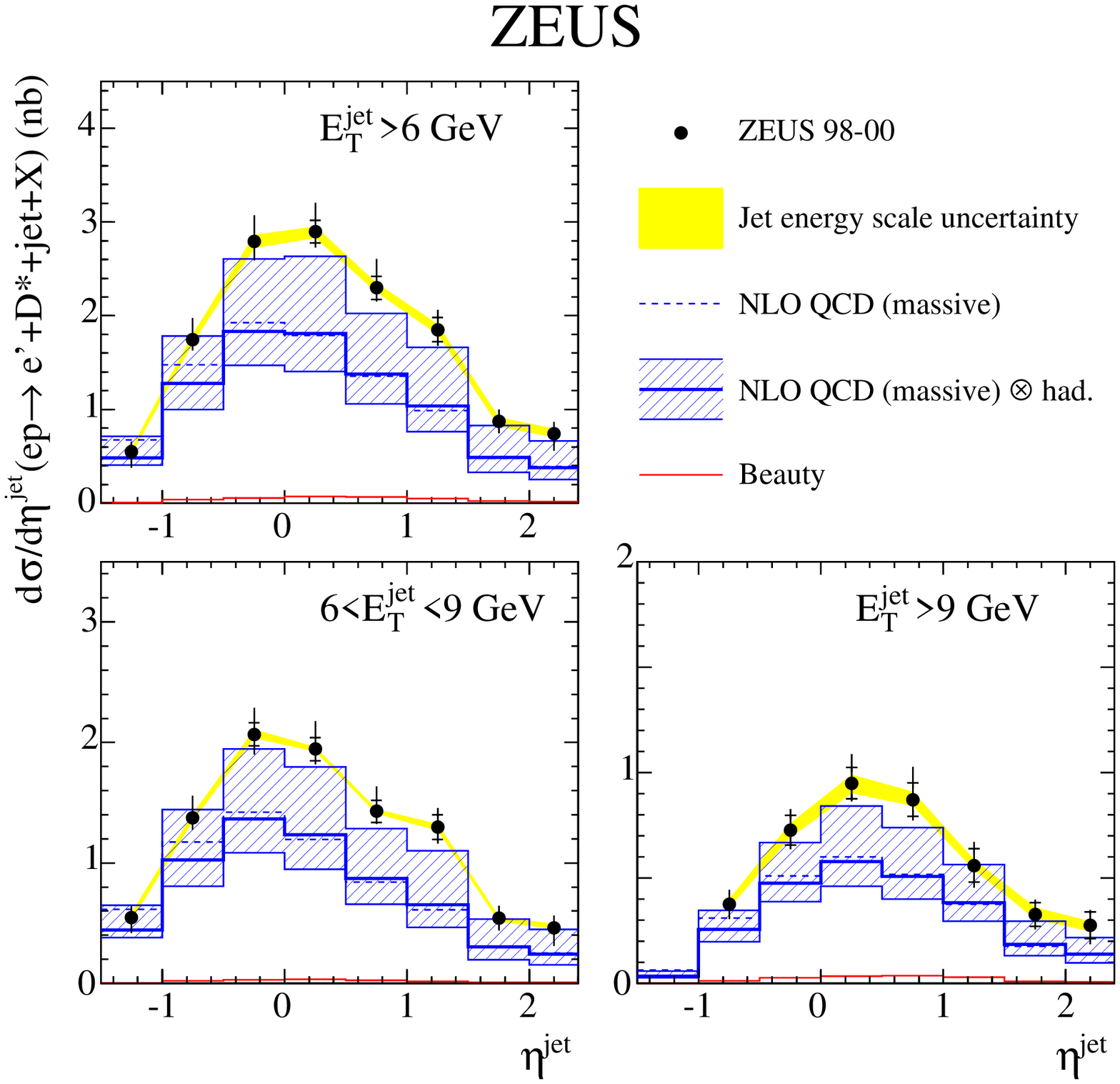}
  \end{center}
  \caption{
Cross-section $d\sigma / d \eta^{\rm jet}$ for jets in events (dots) containing 
at least one $D^*$ meson for different regions in $E_T^{\rm jet}$. The comparison 
is made to massive QCD predictions with (solid line) and without (dotted line) 
hadronisation corrections applied. The theoretical uncertainties (hatched band) 
come from the change in scales simultaneously with the change 
in charm mass. The beauty component is also shown (lower histogram).
  }
  \label{fig-eta}
  \vfill
\end{figure}

\begin{figure}[p]
  \vfill
  \begin{center}
    \includegraphics[width=\linewidth]{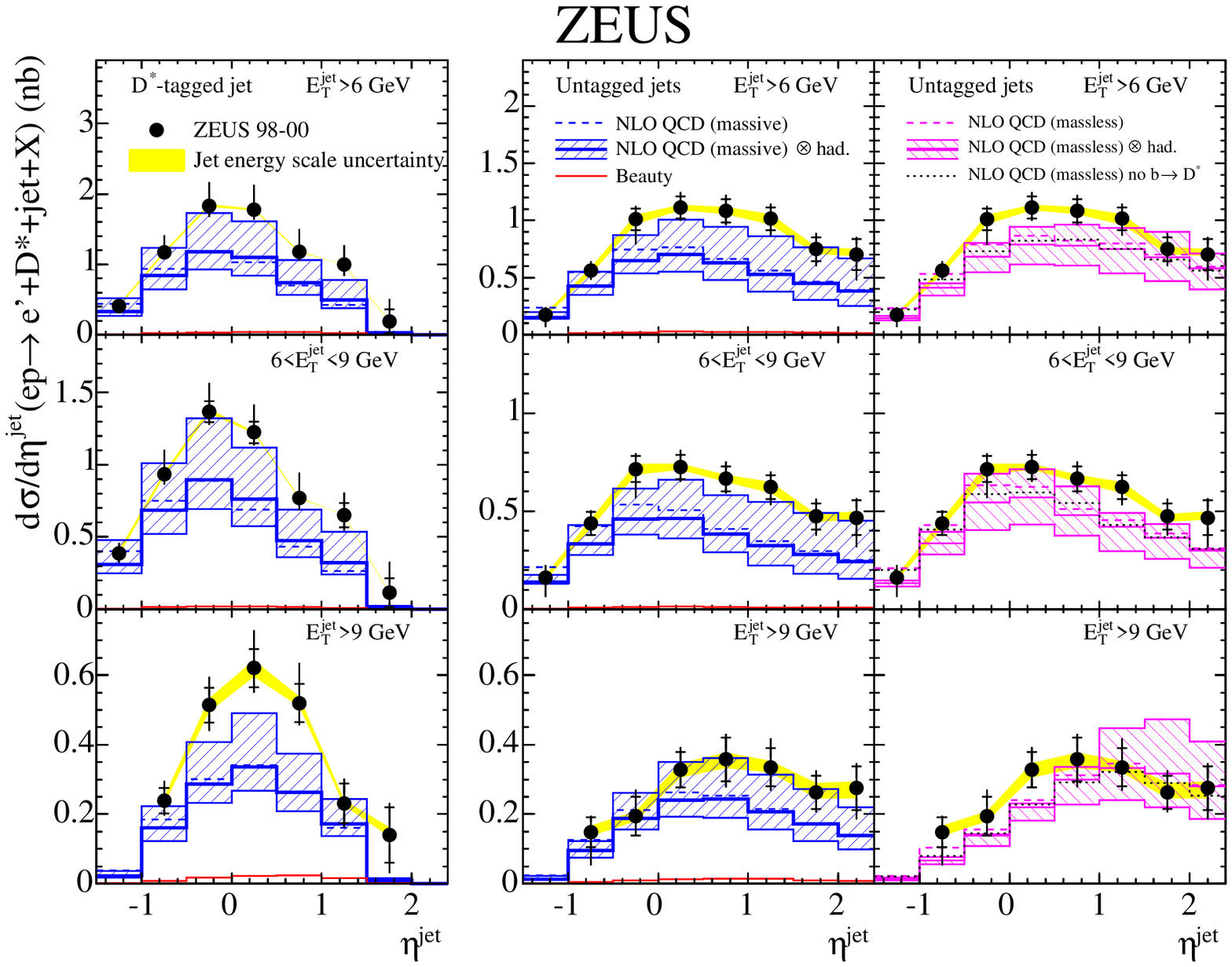}
  \end{center}
  \caption{
Cross-section $d\sigma / d \eta^{\rm jet}$ for $D^*$-tagged jets and untagged jets 
(dots). The comparison is made to massive QCD predictions with (solid line) and without 
(dashed line) hadronisation corrections applied. The 
beauty component is also shown (lower histogram). For the untagged jet distribution, the 
massless QCD predictions are also shown with (solid line) and without (dashed line) 
hadronisation corrections applied.
The theoretical uncertainties (hatched bands) come, in the case of
the massive calculations, from changing renormalisation and
factorisation scales as well as the charm mass simultaneously.
In the case of the massless calculations, they come from changing the
scales only.
The prediction with no component from $b$-quark  fragmentation to a $D^*$ is also shown (dotted line).
  }
  \label{fig-eta_dstarother}
  \vfill
\end{figure}

\begin{figure}[p]
  \begin{center}
    \includegraphics[width=13cm]{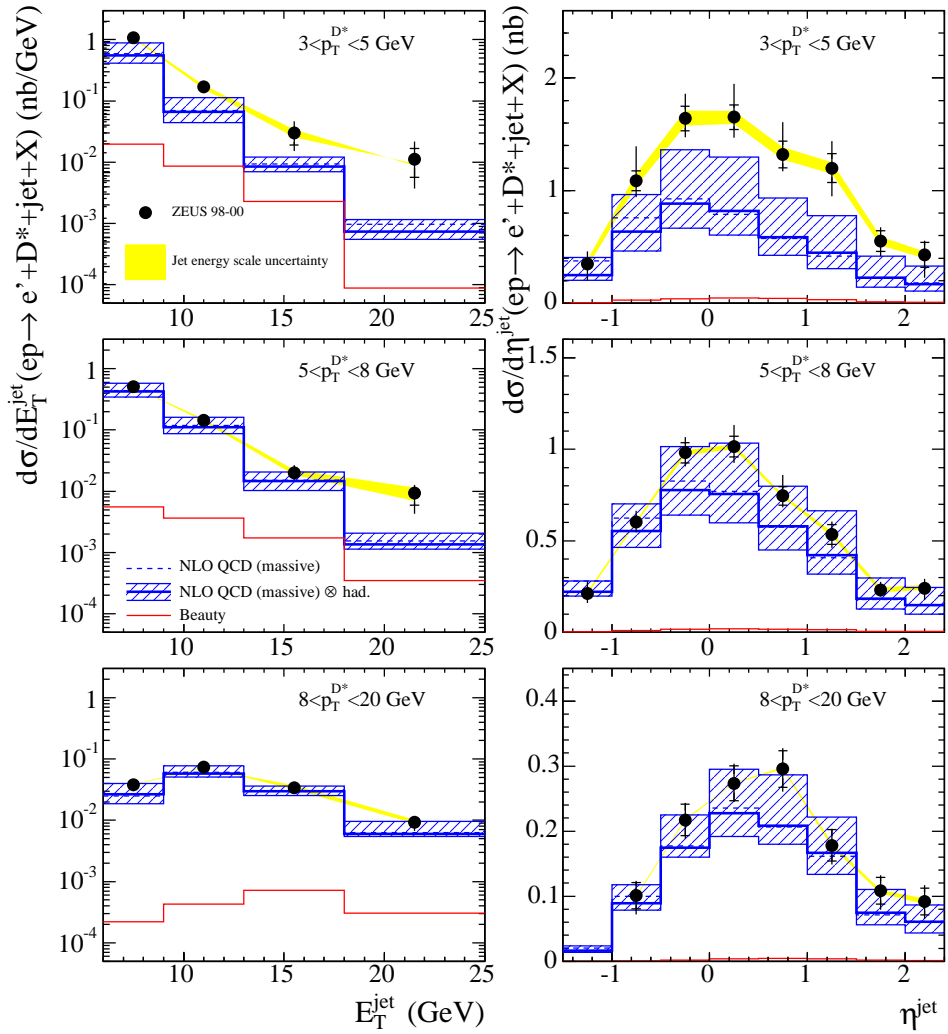}
  \end{center}
  \caption{
Cross-sections $d\sigma/dE_{T}^{\rm jet}$ and $d\sigma/d\eta^{\rm jet}$ in bins of 
$p_{T}^{D^*}$. The data (solid dots) are compared to the massive QCD predictions with 
(solid line) and without (dotted line) hadronisation corrections applied. The 
theoretical uncertainties (hatched band) come from the change in scales simultaneously 
with the change in charm mass. The beauty component is also shown (lower histogram).}
  \label{fig-inptbins}
\end{figure}


\begin{figure}[p]
  \begin{center}
    \includegraphics[width=\linewidth]{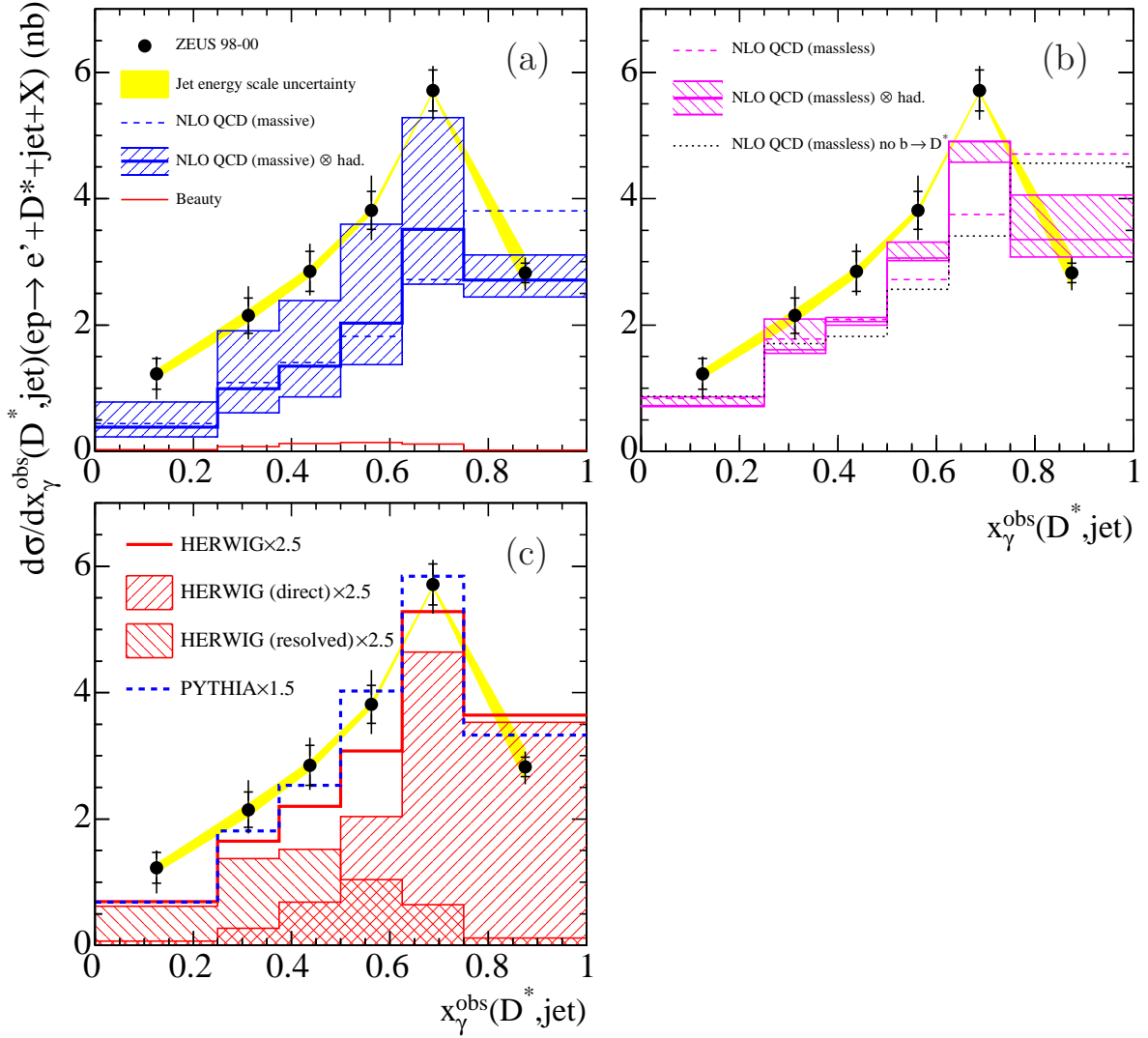}
    \put(-255,390){\makebox(0,0)[tl]{\large (a)}}
    \put(-45,390){\makebox(0,0)[tl]{\large (b)}}
    \put(-255,196){\makebox(0,0)[tl]{\large (c)}}
  \end{center}
  \caption{
Cross-section $d\sigma/dx^{\rm obs}_{\gamma}(D^*,{\rm jet})$ for the events containing a $D^*$ 
meson not associated with a jet. The data (solid dots) are compared to (a) the massive QCD 
predictions with (solid line) and without (dotted line) hadronisation corrections applied. 
The beauty component is also shown (lower histogram). In (b) 
the data are compared to the massless QCD predictions shown with (solid line) and without 
(dotted line) hadronisation corrections applied.
The prediction with no component from $b$-quark fragmentation to a $D^*$ is also shown (dotted line).
The theoretical uncertainties (hatched bands) come, in the case of
the massive calculations, from changing renormalisation and
factorisation scales as well as the charm mass simultaneously.
In the case of the massless calculations, they come from changing the
scales only.
In (c) the data are compared to {\sc Herwig} (solid line) and {\sc Pythia} (dashed line) MC predictions 
normalised to the data. The predicted {\sc Herwig} direct and resolved contributions are also 
shown.
  } 
  \label{fig-xgamma_dsjet}
\end{figure}

\begin{figure}[p]
  \begin{center}
    \includegraphics[width=\linewidth]{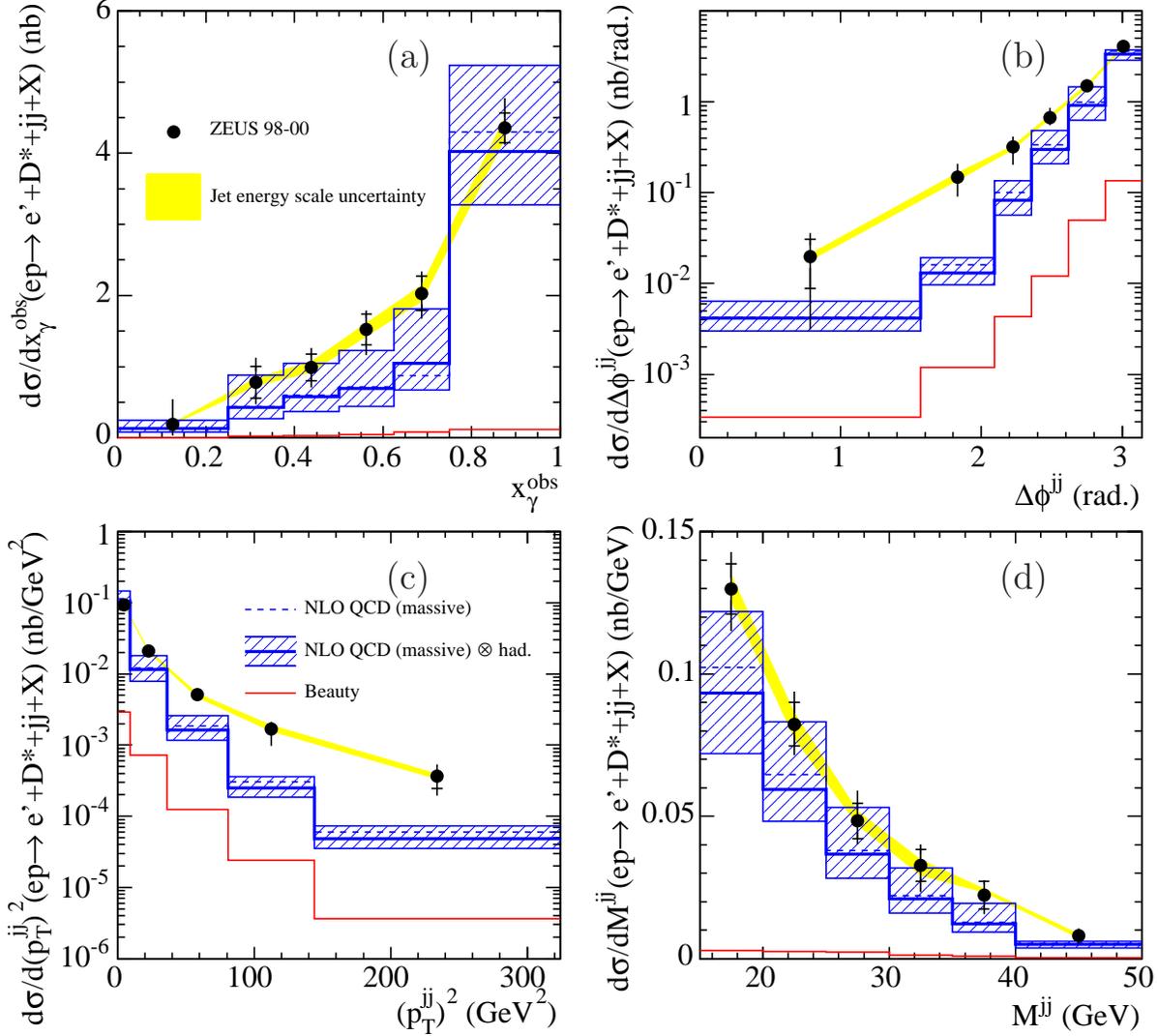}
    \put(-305,390){\makebox(0,0)[tl]{\large (a)}}
    \put(-65,390){\makebox(0,0)[tl]{\large (b)}}
    \put(-305,186){\makebox(0,0)[tl]{\large (c)}}
    \put(-65,186){\makebox(0,0)[tl]{\large (d)}}
  \end{center}
  \caption{
Dijet cross-sections (a) $d\sigma/dx_{\gamma}^{\rm obs}$, (b) $d\sigma/d\Delta\phi^{\rm jj}$, 
(c) $d\sigma/d(p_{T}^{\rm jj})^2$ and (d) $d\sigma/dM^{\rm jj}$ for the data (solid dots) compared to 
massive QCD predictions with (solid line) and without (dotted line) hadronisation corrections 
applied. The theoretical uncertainties (hatched band) come from the change in scales 
simultaneously with the change in charm mass. The beauty component is also shown (lower 
histogram).}
  \label{fig-xgamma}
\end{figure}

\begin{figure}[p]
  \begin{center}
    \includegraphics[width=\linewidth]{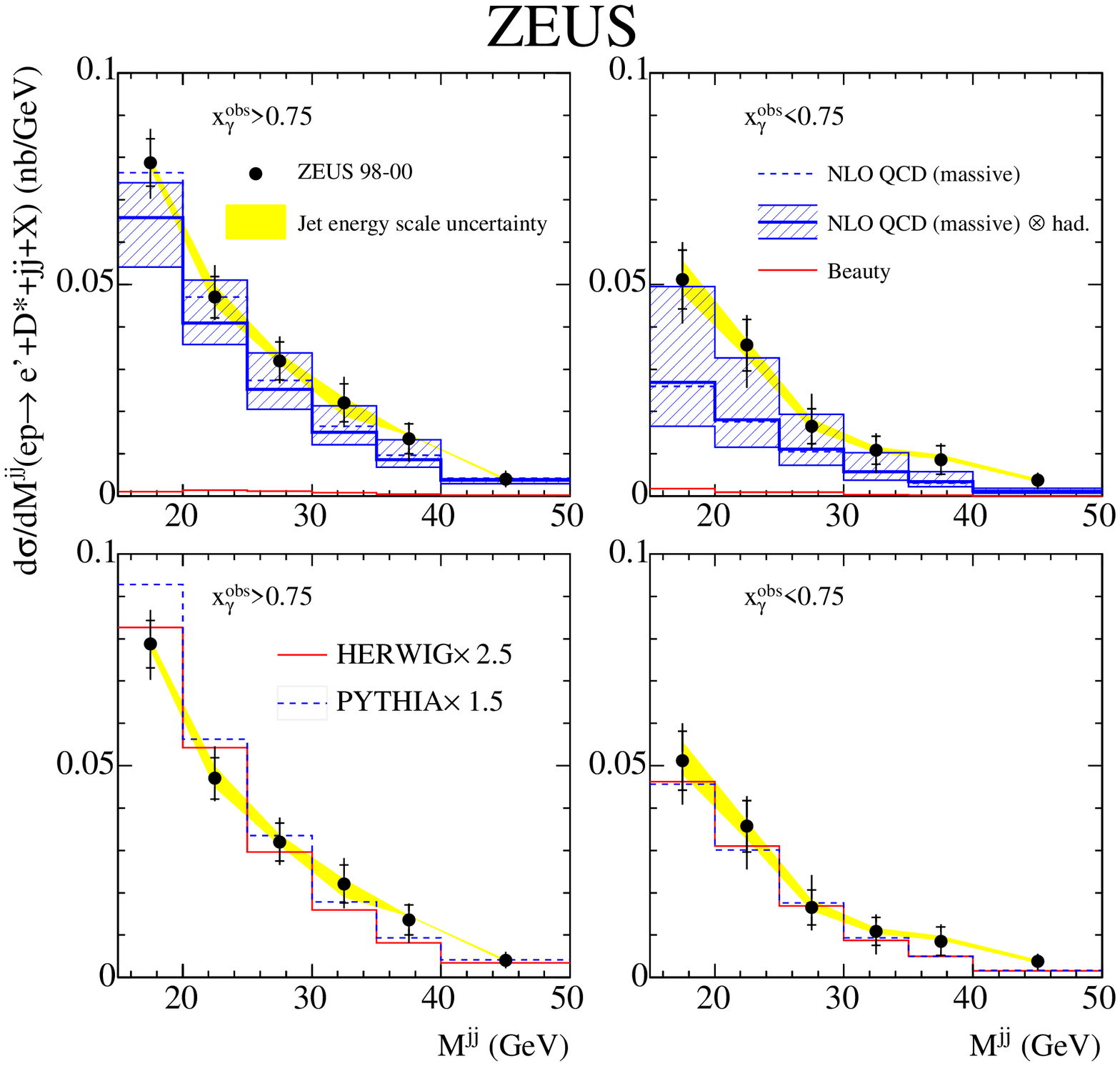}
    \put(-255,390){\makebox(0,0)[tl]{\large (a)}}
    \put(-45,390){\makebox(0,0)[tl]{\large (b)}}
    \put(-255,196){\makebox(0,0)[tl]{\large (c)}}
    \put(-45,196){\makebox(0,0)[tl]{\large (d)}}
  \end{center}
  \caption{
Cross-section $d\sigma/dM^{\rm jj}$ separated into (a,c) direct enriched 
($x_{\gamma}^{\rm obs} > 0.75$) and (b,d) resolved enriched ($x_{\gamma}^{\rm obs} < 0.75$). 
The data (solid dots) are compared (a,b) to the massive QCD prediction with (solid line) 
and without (dotted line) hadronisation corrections applied. The theoretical uncertainties (hatched 
band) come from the change in scales simultaneously with the change in charm mass. The beauty 
component is also shown (lower histogram). The data are also compared (c,d) with {\sc Herwig} 
(solid line) and {\sc Pythia} (dashed line) MC predictions multiplied by the indicated factors. 
}
  \label{fig-mjj}
\end{figure}

\begin{figure}[p]
  \begin{center}
    \includegraphics[width=\linewidth]{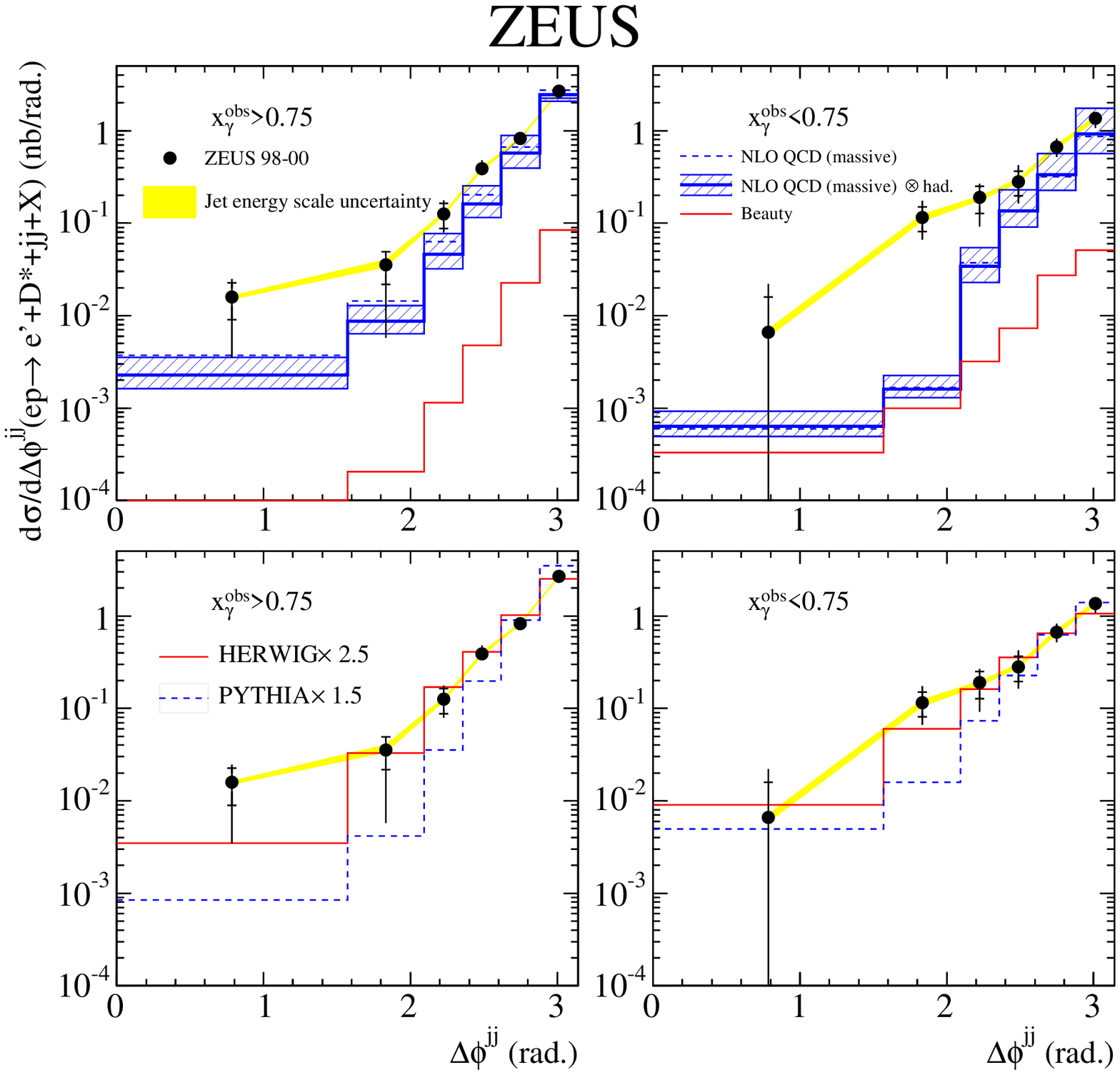}
    \put(-305,390){\makebox(0,0)[tl]{\large (a)}}
    \put(-65,390){\makebox(0,0)[tl]{\large (b)}}
    \put(-305,196){\makebox(0,0)[tl]{\large (c)}}
    \put(-65,196){\makebox(0,0)[tl]{\large (d)}}
  \end{center}
  \caption{
Cross-section $d\sigma/d\Delta\phi^{\rm jj}$ separated into (a,c) direct enriched 
($x_{\gamma}^{\rm obs} > 0.75$) and (b,d) resolved enriched ($x_{\gamma}^{\rm obs} < 0.75$). 
The data (solid dots) are compared (a,b) to the massive QCD prediction with (solid line) 
and without (dotted line) hadronisation corrections applied. The theoretical uncertainties (hatched 
band) come from the change in scales simultaneously with the change in charm mass. The beauty 
component is also shown (lower histogram). The data are also compared (c,d) with {\sc Herwig} 
(solid line) and {\sc Pythia} (dashed line) MC predictions multiplied by the indicated factors. 
}
  \label{fig-dphi}
\end{figure}

\begin{figure}[p]
  \begin{center}
    \includegraphics[width=\linewidth]{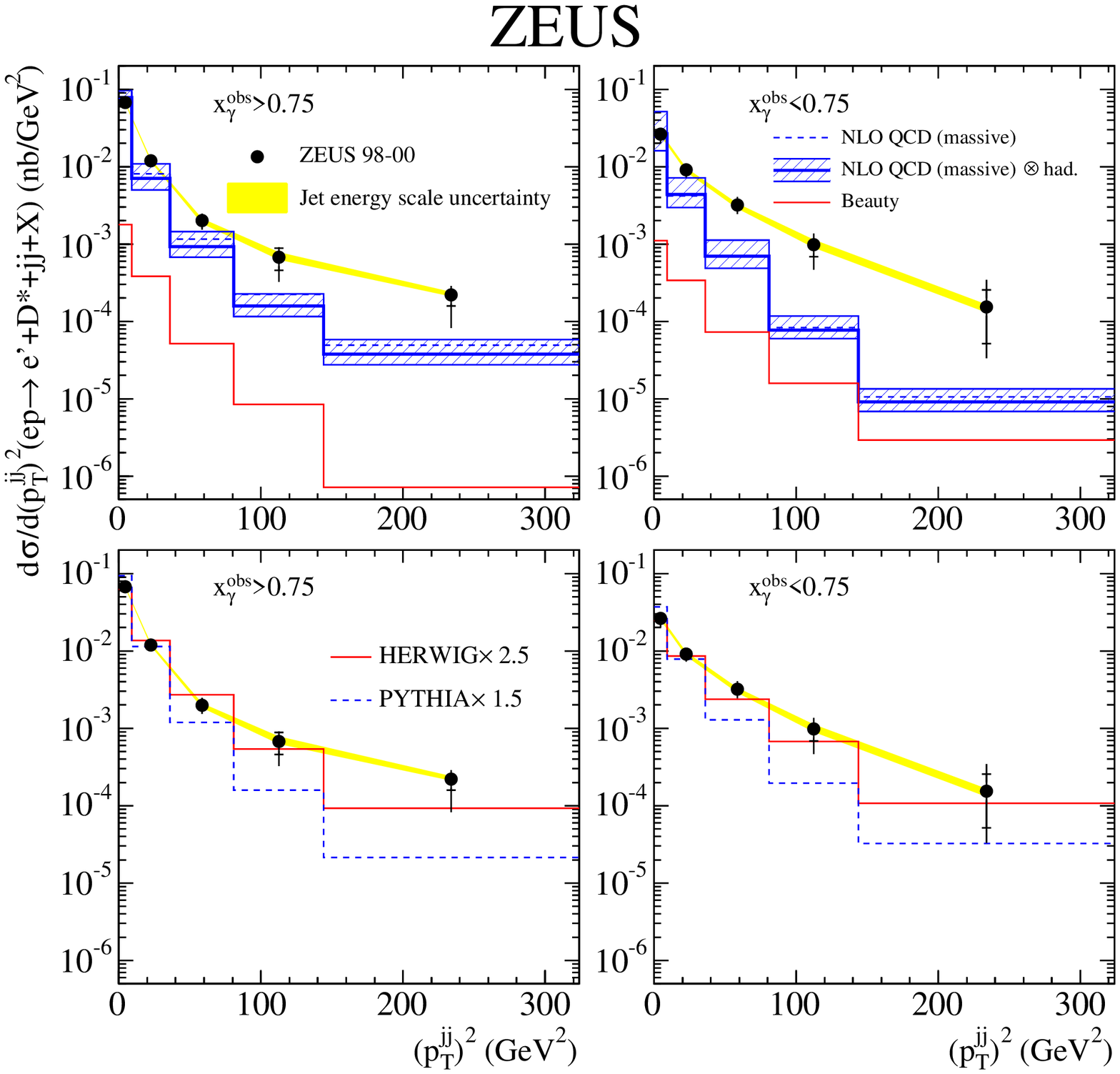}
    \put(-255,390){\makebox(0,0)[tl]{\large (a)}}
    \put(-45,390){\makebox(0,0)[tl]{\large (b)}}
    \put(-255,196){\makebox(0,0)[tl]{\large (c)}}
    \put(-45,196){\makebox(0,0)[tl]{\large (d)}}
  \end{center}
  \caption{
Cross-section $d\sigma/d(p_T^{\rm jj})^2$ separated into (a,c) direct enriched 
($x_{\gamma}^{\rm obs} > 0.75$) and (b,d) resolved enriched ($x_{\gamma}^{\rm obs} < 0.75$). 
The data (solid dots) are compared (a,b) to the massive QCD prediction with (solid line) 
and without (dotted line) hadronisation corrections applied. The theoretical uncertainties (hatched 
band) come from the change in scales simultaneously with the change in charm mass. The beauty 
component is also shown (lower histogram). The data are also compared (c,d) with {\sc Herwig} 
(solid line) and {\sc Pythia} (dashed line) MC predictions multiplied by the indicated factors. 
}
  \label{fig-ptjj}
\end{figure}
}

\end{document}